\documentclass[12pt]{extarticle}
\pdfoutput=1

\usepackage{mathtools}
\usepackage{amssymb}
\usepackage{braket}
\usepackage{tensor}
\usepackage{slashed}
\usepackage{dsfont}
\usepackage{graphicx}
\usepackage{array}
\usepackage{float}
\usepackage[T1]{fontenc}
\usepackage{xcolor}
\usepackage{enumitem}
\usepackage{setspace}
\usepackage{geometry}
\usepackage{tikz}
\usetikzlibrary{
    shapes.callouts,
    positioning,
    calc
}
\usepackage[
    colorlinks=true,
    linkcolor=blue,
    citecolor=red,
    urlcolor=blue
]{hyperref}
\numberwithin{equation}{section}
\allowdisplaybreaks[2]

\newcommand{\tr}{\operatorname{tr}}
\newcommand{\pd}{\partial}

\newcommand{\qqv}{Q^{\vee}}

\newcommand{\bz}{\mathbb{Z}}

\title{Localization and Abelianization of Strings \\on Group Manifolds: \\ The Non-simply Connected Case}

\makeatletter
\renewcommand\@date{
  \vspace{-\baselineskip}
  \large\centering
  Yongchao L\"u
 
  \normalsize
  lychaoaa@gmail.com
}
\makeatother

\begin{document}
\maketitle
\begin{abstract}
We study the partition functions of Wess--Zumino--Witten (WZW) models with compact connected simple Lie group manifolds that are not simply connected. Starting from the modular-invariant partition function of Felder--Gaw\k{e}dzki--Kupiainen (FGK), we derive a localization formula, which can also be obtained directly by supersymmetric localization of the corresponding supersymmetric WZW model. Our results reveal a variety of topological effects associated with the nontrivial topology of the target group. In particular, the Wess--Zumino amplitude is governed by FGK cocycles, while the fermion Pfaffians exhibit global anomalies associated with the holonomies of Pfaffian line bundles, captured by relative Rochlin invariants. Together, these results provide a unified description of the topological contributions to the semiclassical localization of WZW models with non-simply connected target groups.
\end{abstract}
\tableofcontents

\section{Introduction}

The Wess–Zumino–Witten (WZW) model is a distinguished string background whose exact solvability has provided deep insights into both physics and mathematics~\cite{Witten1983WZW}. Its partition function can be computed exactly using the representation theory of affine Kac–Moody algebras, or evaluated semiclassically by methods such as supersymmetric localization~\cite{murthyW25}. 
In a companion work~\cite{Lu_wzwlocalization2026}, we studied the partition function of the WZW model for compact, connected, and \emph{simply connected} simple Lie groups using supersymmetric localization of the path integral of the corresponding supersymmetric WZW (SWZW) model~\cite{DiVecchiaKnizhnikRossi1984SWZW}. A central step in the localization computation is the abelianization of the localizing solutions, which reduces the non-abelian path integral to a sum over winding sectors on the maximal torus. We showed that the Wess–Zumino (WZ) amplitude produces an additional factor of topological origin~\cite{Wesszumino1971}. This factor is essential for establishing agreement between the localization result and the Hamiltonian formalism.

The purpose of the present work is to extend this analysis to compact connected simple Lie groups that are not simply connected. The global topology of the target group manifold modifies the set of topological sectors contributing to the path integral. Furthermore, maps from the worldsheet into the target group manifold can carry nontrivial topological information associated with the global form of the Lie group. Correspondingly, the abelianized localizing locus is no longer characterized solely by the winding sectors familiar from the simply connected case. Additional discrete topological data associated with the nontrivial fundamental group of the target space must also be taken into account.

A useful starting point is the modular-invariant partition function constructed by Felder--Gaw\k{e}dzki--Kupiainen (FGK) for non-simply connected simple Lie groups~\cite{FGK1988}. It can be understood as a simple-current orbifold of the diagonal modular invariant associated with the simply connected covering group~\cite{SchellekensYankielowicz1989Extended, KreuzerSchellekens1993}. Similar to the simply connected case, our task is to rewrite the FGK modular invariant in terms of generalized Siegel--Narain theta functions. The novel aspect arises from the nontrivial phases associated with the simple-current orbifold action.  As a result, the relevant lattices and their shifts are modified by the global form of the Lie group, and additional phases appear.

The new topological effects arise from several interrelated sources. First, the
WZ amplitude is sensitive to the global topology of the target group manifold.
In the presence of nontrivial topological sectors, it requires a global
refinement encoded by a cocycle phase factor, as identified by Felder,
Gaw\k{e}dzki, and Kupiainen~\cite{FGK1988}. This cocycle reflects the global
gerbe data underlying the WZ amplitude~\cite{GawedzkiWaldorf2009PWMulGerb} and, after abelianization, is naturally
related to the global topological data of the $B$-field. Second, the Pfaffian
for Majorana--Weyl fermions may acquire nontrivial holonomy under transport
around noncontractible loops in the moduli space of flat background gauge
configurations. Consequently, the Pfaffian line bundle need not be globally
trivial, and its holonomy contributes a global anomaly phase to the path
integral~\cite{Witten1985Global}, characterized by the holonomy of the relative
Pfaffian line, which may be expressed through mapping-torus eta invariants and,
on the torus, relative Rochlin invariants~\cite{LeeMillerWeintraub1988}. These
additional phase factors have no counterpart in the localization formula for
simply connected target groups.

The relation between these two types of global data has appeared in several
closely related contexts. Global anomalies of Majorana--Weyl fermions in
two-dimensional sigma models and their relation to the WZ term have recently
been studied in~\cite{Choi:2025MajoWeyl}. The interplay between the global
phase of the fermionic Pfaffian and the $B$-field/WZ amplitude is familiar from
the global formulation of the worldsheet path integral~\cite{Witten1999Worldsheet}.
More generally, global anomalies of fermions and the global topological data of
$B$-fields have recently been studied together from a bordism
perspective~\cite{SaitoTachikawa:2025GS}.

Taken together, these WZ and fermionic global effects significantly enrich the spectrum of topological phenomena that arise in the semiclassical treatment of WZW models with compact connected simple Lie groups.  In the localization computation, these effects manifest directly through the classical action evaluated on the localizing solutions and through the corresponding one-loop determinants. In this way, the localization formula provides a semiclassical realization of the global topological data captured by the Hamiltonian description.

Our main result is a localization formula for the SWZW partition function with a general compact connected simple Lie group. The formula organizes the contributions according to the relevant topological sectors and makes explicit the phases arising from both the fermion global anomaly and the WZ amplitude, schematically given by
\begin{align}
    Z_{\rm SWZW}
=
\sum_{\sigma\in W} 
\sum_{m,w\in\Lambda^\vee}
\underbrace{\phi_{[m],[w]}}_{\text{FGK}}
\underbrace{\nu_{[m],[w]}}_{\text{Rochlin}}
\,
\mathcal{F}_{m,w}^\sigma.
\end{align}
Here $\sigma$, $m$, and $w$ label the localizing solutions through the Weyl group and the winding sectors on the torus, with $[m]$ and $[w]$ denoting the corresponding homotopy classes of loops in the Lie group. The phase $\phi$ is the FGK cocycle associated with the WZ amplitude, while $\nu$ encodes the relative Rochlin invariants for the fermion global anomaly. The quantity $\mathcal{F}_{m,w}^{\sigma}$ contains the remaining localization contribution, including the classical action and one-loop determinants, and has the same structure as in the simply connected case. The resulting expression agrees with the Hamiltonian derivation, demonstrating that the same global topological data appear consistently in both the canonical and semiclassical approaches.

One of the essential features of the localization is the abelianization. The global topology of the target group induces a simple--current orbifold structure on the Narain lattice description. Consequently, the SWZW partition function can be expressed as an orbifold sum of generalized Siegel--Narain theta functions, together with discrete torsion arising from the global fermion anomaly. Semiclassically, in light of the relation between the WZ amplitude and the amplitude of a constant Kalb–Ramond (KR) $B$--field, the FGK cocycle naturally emerges in the Narain description, encoded in the generalized Siegel–Narain theta functions through a lattice cocycle in the winding sector of the torus. This offers an explicit account of how the global topological data manifest in terms of the lattice data of the abelianized description. 

An instructive feature emerges upon circle reduction. Although the FGK cocycle becomes trivial in the resulting one--dimensional description, the global anomaly of the fermions persists. In the generalized Frenkel formula, this surviving effect appears as a non--trivial phase determined by the global anomaly of one--dimensional fermions, naturally characterized as the holonomy of the fermion Pfaffian line bundle, equivalently by a mod-$2$ index on the associated mapping torus. Thus, the circle reduction separates the two topological effects that are intertwined in the two-dimensional theory: the FGK cocycle is trivialized through dimensional reduction, while the fermion global anomaly survives as a genuine phase of the reduced theory. This provides a coherent picture of how the global topology of the group manifold is transmitted from the two-dimensional WZW model to its one-dimensional reduction.

The paper is organized as follows. In Section~\ref{sec:hamiltonian}, we derive the localization formula starting from the FGK modular invariant of the WZW model. In Section~\ref{sec:localization}, we analyze the topological aspects of the WZ amplitude and determine the associated FGK cocycle, as well as the global anomaly of the fermion Pfaffian in terms of the relative Rochlin invariant. In Section~\ref{sec:narain}, we explore the manifestation of the FGK cocycle in the Kalb--Ramond $B$-field amplitude and express the SWZW partition function in terms of generalized Siegel--Narain theta functions. In Section~\ref{sec:frenkel}, we consider the particle limit and recover the generalized Frenkel formula, providing a global-anomaly interpretation of the phase factor. Finally, in Section~\ref{sec:summary}, we conclude by discussing our results, their implications, and future directions. 
Several technical details are collected in the appendices. Appendix~\ref{apx:lieGroup} reviews our conventions for Lie groups, global forms, and the associated lattices. Appendix~\ref{apx:constraints} summarizes the WZW and Chern--Simons level constraints, while Appendix~\ref{apx:phasePhi} gives the explicit form of the new phase factors entering the localization formula. Appendix~\ref{apx:phaseNu} analyzes the case-by-case triviality conditions for the fermion global anomaly phase. Appendix~\ref{apx:arfRochlin} reviews the relation between quadratic refinements, Arf invariants, and relative Rochlin invariants underlying the fermion global anomaly phase. Finally, Appendix~\ref{apx:SiegelNarain} defines the generalized Siegel--Narain theta functions and summarizes the properties used in the main text.

\section{Localization Formula from FGK Modular Invariants} 
\label{sec:hamiltonian}

For a WZW model with a non-simply connected compact simple Lie group, we derive a localization formula starting from the modular invariant partition function of Felder, Gaw\k{e}dzki, and Kupiainen (FGK). The resulting formula refines the non-simply connected localization formula of Murthy and Witten by revealing additional discrete phases that encode the nontrivial topology of the target group manifold. 
Throughout this work, we adopt the group-theoretic notation collected in Appendix~\ref{apx:lieGroup}.

\subsection{The FGK torus partition function}

We consider the WZW model with compact, connected, simple Lie group $G = \tilde G/ \mathcal{C}$. The model is further specified by a level $k$ which is restricted by
the WZW and Chern--Simons (CS) constraints summarized in Appendix~\ref{apx:constraints}.  The spectrum can be obtained either geometrically, through quantization of the centrally extended loop group of
$G$~\cite{FGK1988}, or equivalently as a simple-current orbifold by
$\mathcal C$ of the WZW model for the universal cover $\widetilde G$~\cite{KreuzerSchellekens1993}.
The latter has the \emph{diagonal} spectrum $(\lambda,\lambda)$ with
$\lambda\in P_+^k$. For $\omega\in\mathcal C$, we choose the corresponding minuscule
coweight representative, which we denote by the same symbol $\omega$.
The associated symmetry of the extended Dynkin diagram determines
$\sigma_\omega\in W$ with determinant
\begin{align}
\label{eq:WeylSignSimpleCurrent}
\det(\sigma_\omega)
= (-1)^{2(\rho,\omega)}
= (-1)^{h^\vee\omega^2}.
\end{align}
The corresponding simple current acts as
\begin{align}
J_\omega(\lambda)
=
\sigma_\omega(\lambda)+k\omega.
\end{align}
The spectrum of the WZW model with target group $G$ is then encoded in the FGK mass matrix $M_{\lambda,\lambda'}$, which specifies the \emph{non-diagonal} pairing of left- and right-moving integrable representations~\cite{FGK1988}, 
\begin{align}
M_{\lambda,\lambda'}
& =  \sum_{\omega\in\mathcal C} 
\delta_{\mathcal{C}}(
\lambda +\frac{k}{2}\omega
+\frac{1}{2}\eta^\epsilon_\omega)
\delta_{\,J_\omega(\lambda),\, \lambda'} 
\\
& =
\frac{1}{|\mathcal C|}
\sum_{\mu,\omega\in\mathcal C}
(-1)^{\epsilon \mu \wedge
 \omega} e^{2\pi i(\lambda,\mu) + \pi i  k(\mu,\omega)}
\delta_{\, \sigma_\omega(\lambda)+k\omega,\,\lambda'},
\end{align}
where $\delta_{\mathcal C}(x)=1$ if $(x,\omega)\in\mathbb Z$ for all
$\omega\in\mathcal C$, and vanishes otherwise. Nontrivial discrete torsion occurs only for $\mathfrak g=D_{2n}$ with $\mathcal C=(\mathbb Z_2)^2$, where it is parametrized by $\epsilon\in\{0,1\}$ and
$(\eta^\epsilon_\omega,\mu)
=
\epsilon\,\omega\wedge\mu$. We also write $(-1)^{\epsilon \mu \wedge
 \omega}$ as $(\pm 1)^{\mu \wedge
 \omega}$ for convenience. 
Our conventions for $\omega\wedge\mu$ and further group-theoretic details are collected in Appendix~\ref{apx:lieGroup}.

The torus partition function is given by the FGK modular invariant, which is a non-diagonal combination of affine Kac–Moody characters: 
\begin{align}
Z_{\text{WZW}}
=
C_k
\sum_{\lambda,\lambda’ \in P_+^k}
M_{\lambda,\lambda’}
\chi_{\lambda,k}(\tau,u)
\overline{\chi_{\lambda’,k}(\tau,v)} ,
\end{align}
where affine Kac–Moody characters are given by the Weyl--Kac formula
\begin{align}
    \chi_{\lambda,k} = \frac{N_{\lambda+\rho, \kappa}}{N_{\rho,h^\vee}},\quad N_{\lambda,\kappa}
=
\sum_{\sigma\in W}
\det(\sigma)\,
\Theta_{\sigma(\lambda),\kappa}, 
\end{align}
and the non-holomorphic factor for cancellation of the equivariant modular anomaly is
\begin{align}
C_k = \exp\{ \frac{\pi ik}{\tau_2}\left((u, \mathrm{Im}\,u ) + (\bar{v}, \mathrm{Im}\, \bar v )\right)\}.
 \end{align}
Here $\tau$ denotes the modular parameter of the torus, while $u$ and $v$ are Cartan-valued chemical potentials for the $G_L \times G_R$ symmetry, respectively.

Adding Majorana fermions in the adjoint representation yields an $\mathcal N=(1,1)$ supersymmetric WZW (SWZW) model. We impose periodic boundary conditions for the fermions. In the presence of a Cartan-valued fermion zero-mode insertion, which renders the fermion trace non-vanishing, the partition function of the adjoint Majorana fermions is
\begin{align}
Z_{\mathrm{F}}
=
2^r\, C_{h^\vee}
N_{\rho,h^\vee}
\bar{N}_{\rho,h^\vee},
\end{align}
where
\begin{align*}
    C_{h^\vee} = \exp\{ \frac{\pi ih^\vee}{\tau_2}\left((u, \mathrm{Im}\,u ) + (\bar{v}, \mathrm{Im}\,\bar v )\right)\},
\end{align*}
and the factor $2^r\,$ with $r = \text{rank}(G)$ arises from the quantization of the Cartan--valued Majorana zero modes.

Since the bosonic and fermionic sectors decouple, the SWZW partition function factorizes and gives
\begin{align}
Z_{\mathrm{SWZW}}
=
Z_{\mathrm{WZW}} Z_{\mathrm{F}} = 2^r\, C_\kappa N_{\rho,h^\vee}
\bar{N}_{\rho,h^\vee} \sum_{\lambda,\lambda’ \in P_+^k}
M_{\lambda,\lambda’}
\chi_{\lambda,k}
\bar{\chi}_{\lambda' ,k} .
\end{align}
where $\kappa=k+h^\vee$ and the non-holomorphic prefactors combine as
\begin{align}
    C_{\kappa} = C_k C_{h^\vee} = \exp\{ \frac{\pi i \kappa}{\tau_2}\left((u, \mathrm{Im}\,u ) + (\bar{v}, \mathrm{Im}\,\bar v )\right)\}.
\end{align}

The fermion contribution cancels the Weyl–Kac numerator factors in $Z_{\mathrm{WZW}}$. Furthermore, the shift $\lambda\mapsto\lambda+\rho$ maps $P_+^k$ to $P_{++}^\kappa$. Since $N_{\lambda,\kappa}$ vanishes for non-regular $\lambda$, the sum may then be extended to all of $P_+^\kappa$. To account for the simple-current action under this shift, we note that $\rho-\sigma_\omega(\rho) = h^\vee \omega$, which lifts $J_\omega$ to the shifted action $\widehat J_\omega$:
\begin{align}
\widehat J_\omega (\lambda) \equiv J_{\omega}(\lambda -\rho) + \rho = J_{\omega}(\lambda) + h^\vee \omega = \sigma_\omega(\lambda) + \kappa \omega, \quad \lambda \in P_+^\kappa.
\end{align}
Accordingly we define the supersymmetric mass matrix as
\begin{align}
\widehat{M}_{\lambda,\lambda’}
& = \sum_{\omega \in \mathcal{C}} 
\delta_{\mathcal{C}}\!\left(
\lambda -\rho
+\frac{k}{2}\omega
+\frac{1}{2}\eta^{\epsilon}_\omega
\right)
\delta_{
\widehat{J}_\omega(\lambda),\, \lambda'} \\
& = 
\frac{1}{\lvert \mathcal{C} \rvert}
\sum_{\mu,\omega \in \mathcal{C}}
  (-1)^{\epsilon \mu \wedge
 \omega} (-1)^{h^\vee \mu^2} e^{2\pi i(\lambda,\mu) + \pi i  k(\mu,\omega)}
\delta_{
s_\omega(\lambda)+\kappa\omega, \,\lambda'},
\end{align}
Thus
\begin{align}
Z_{\mathrm{SWZW}} =
2^r\, C_\kappa 
\sum_{\lambda,\lambda’ \in P_+^\kappa}
\widehat{M}_{\lambda,\lambda’}
N_{\lambda,\kappa}(\tau, u)
\overline{N_{\lambda’,\kappa}(\tau, v)},
\end{align}

Using~\eqref{eq:WeylSignSimpleCurrent},
\begin{align}
N_{\lambda,\kappa}
=
(-1)^{h^\vee\omega^2}
N_{\sigma_\omega(\lambda),\kappa}.
\end{align}
Moreover, $e^{2\pi i(\lambda,\mu)}$ is Weyl invariant since
$\sigma(\lambda)-\lambda\in Q$ and $(Q,P^\vee)\subset\mathbb Z$.
We may therefore absorb the Weyl sign and rewrite the partition
function as
\begin{align}
Z_{\rm SWZW}
&=
2^r\, C_{\kappa} \frac{1}{\lvert \mathcal{C} \rvert}
\sum_{\mu,\omega \in \mathcal{C}}
\Phi_{\mu,\omega}
\sum_{\lambda \in \sigma_\omega(P_+^\kappa)}
e^{2\pi i(\lambda,\mu)}
N_{\lambda,\kappa}
\bar{N}_{\lambda+\kappa\omega,\kappa},
\end{align}
where we define 
\begin{align} \label{eq:Phi}
\Phi_{\mu,\omega}
=
(-1)^{\epsilon \mu \wedge
 \omega} (-1)^{h^\vee (\mu^2+\omega^2) } e^{ \pi i  k(\mu,\omega)}.
\end{align}
By expanding $\bar{N}_{\lambda+\kappa\omega,\kappa}$ and noting that $\sigma(\mu) - \mu \in Q^\vee$ for $\sigma \in W$, $\mu \in P^\vee$, and $\bar\Theta_{\lambda + \kappa\gamma, \kappa} = \bar\Theta_{\lambda, \kappa}$ for $\gamma \in Q^\vee$, we can unfold the summation range of $\lambda$ to $P/\kappa Q^\vee$:
\begin{align}
    Z_{\rm SWZW}=
2^r\, C_{\kappa} \frac{1}{\lvert \mathcal{C} \rvert}
\sum_{\mu,\omega \in \mathcal{C}}
\Phi_{\mu,\omega}
\sum_{\lambda \in P/\kappa \qqv}
e^{2\pi i(\lambda,\mu)}
N_{\lambda,\kappa}
\bar{\Theta}_{\lambda+\kappa\omega,\kappa}
\nonumber.
\end{align}
Further expanding $N_{\lambda,\kappa}$ we obtain
\begin{align}
Z_{\rm SWZW}
=
2^r\, C_{\kappa} 
\sum_{\sigma\in W}
\det(\sigma)
\frac{1}{\lvert \mathcal{C} \rvert}
\sum_{\mu,\omega \in \mathcal{C}}
\Phi_{\mu,\omega}
\sum_{\lambda \in P/\kappa \qqv}
e^{2\pi i(\lambda,\mu)}
\Theta_{\sigma(\lambda),\kappa}
\bar{\Theta}_{\lambda+\kappa\omega,\kappa},
\end{align}
Applying $\Theta_{\sigma(\lambda),\kappa}(\tau,u)=\Theta_{\lambda,\kappa}(\tau,\sigma(u))$, we lift the Weyl-group action on the weight label $\lambda$ to an action on the chemical potential. Next, we introduce the theta function combination
\begin{align}
\label{eq:mcZwithT}
\mathcal{Z} = \frac{1}{\lvert \mathcal{C} \rvert}
\sum_{\mu,\omega \in \mathcal{C}}
\Phi_{\mu,\omega} \mathcal{T}_{\mu,\omega},
\end{align}
with 
\begin{align}
\label{eq:TMuOmega}
    \mathcal{T}_{\mu,\omega} = 
2^r\, C_{\kappa}
\sum_{\lambda \in P/\kappa \qqv}
e^{2\pi i(\lambda,\mu)}
\Theta_{\lambda,\kappa}
\bar{\Theta}_{\lambda+\kappa\omega,\kappa}.
\end{align}
This allows us to express the SWZW partition function compactly as
\begin{align}
Z_{\rm SWZW}
= 
\sum_{\sigma\in W}
\det(\sigma) \mathcal{Z}^\sigma,   
\end{align}
with $\mathcal{Z}^\sigma(\tau, u, v) = \mathcal{Z}(\tau, \sigma(u),v)$. 

Here, $\mathcal Z$ encapsulates all the orbifold data. With the
redefinitions
\begin{align}
\label{eq:EandZ}
\mathcal E_{\mu,\omega}
=
e^{-\pi i\kappa(\mu,\omega)}\Phi_{\mu,\omega},\quad
\mathcal Z_{\mu,\omega}
=
e^{\pi i\kappa(\mu,\omega)}\mathcal T_{\mu,\omega},
\end{align}
the resulting factors acquire natural physical interpretations as the
discrete torsion phase and twisted-sector partition function,
respectively, of a shift orbifold by $\mathcal C$. Accordingly,
\begin{align}
\mathcal Z
=
\frac{1}{|\mathcal C|}
\sum_{\mu,\omega\in\mathcal C}
\mathcal E_{\mu,\omega}\,
\mathcal Z_{\mu,\omega}
\end{align}
takes the standard form of an orbifold partition function. In the present setting, this is the $\mathcal C$ orbifold of the supersymmetric Narain system obtained from the localization of the simply connected SWZW model in~\cite{Lu_wzwlocalization2026}. We return to this interpretation in Section~\ref{sec:narain}.

\subsection{The localization formula}
We now convert $\mathcal{T}_{\mu,\omega}$ into the lattice sum appropriate for localization.  Expanding the theta functions 
\begin{align}
    \Theta_{\lambda,\kappa}(\tau,u)
=
\sum_{\gamma\in Q^\vee}
\exp\{\frac{\pi i \tau}{\kappa}(\lambda+\kappa\gamma)^2 + 2 \pi i (\lambda+\kappa\gamma,u)\}
,
\end{align}
and unfolding $P/\kappa Q^\vee$ to $P$ gives
\begin{align}
\mathcal{T}_{\mu,\omega}
= 2^r\, C_{\kappa} 
\sum_{\lambda\in P}
\sum_{\gamma\in \qqv}
e^{2\pi i(\lambda,\mu)}
e^{\frac{\pi i \tau }{\kappa}\lambda^2 -\frac{\pi i \bar\tau }{\kappa}
\left(\lambda+\kappa(\gamma+\omega)\right)^2}
e^{2\pi i (\lambda, u)
-2\pi i (\lambda+\kappa(\gamma+\omega),\bar v)}.
\end{align}
Then separating the $w$-dependent and $\lambda$-dependent factors gives
\begin{align}
\mathcal{T}_{\mu,\omega} = 2^r\, C_{\kappa}
\sum_{w\in \qqv}
g(w +\omega)
\sum_{\lambda\in P}
e^{2\pi i(\lambda,\mu)}
f_{w +\omega}(\lambda),
\end{align}
where
\begin{align}
g(\gamma)
& =
\exp\{-\pi i\bar{\tau}\kappa\gamma^2
-2\pi i\kappa(\gamma,v)\},\\
f_\gamma(\lambda)
& =
\exp \{\frac{\pi i}{\kappa}
(\tau-\bar{\tau})\lambda^2
+2\pi i
\left(
\lambda,
u-\bar{v}-\bar{\tau}\gamma
\right)\}.
\end{align}
The remaining weight-lattice sum is evaluated by Poisson resummation.
Since the lattice dual to $P$ is $Q^\vee$, Poisson resummation gives
\begin{align}
\sum_{\lambda\in P}
e^{2\pi i(\lambda,\mu)}
f_\gamma(\lambda)
=
{\rm Vol}(\qqv)
\sum_{m\in \qqv}
F_\gamma(m+\mu),
\end{align}
where
\begin{align}
F_\gamma(m)
=
\int_{\mathfrak{h}}
e^{2\pi i(x,m)}
f_\gamma(x)\,dx =
\left(
\frac{\kappa}{2\tau_2}
\right)^{\frac{r}{2}}
\exp\{-\frac{\pi\kappa}{2\tau_2}
\left(
m-\bar{\tau}\gamma
+u-\bar{v}
\right)^2\}.
\end{align}
Substituting back we obtain
\begin{align}
\mathcal{T}_{\mu,\omega}
& =
2^r\, C_{\kappa} {\rm Vol}(\qqv)
\left(
\frac{\kappa}{2\tau_2}
\right)^{\frac{r}{2}}
\sum_{w,m\in Q^\vee} 
g(w+\omega)
F_{w+\omega}(m+\mu) \nonumber\\
& = 2^r\,  {\rm Vol}(Q^\vee)
\left(\frac{\kappa}{2 \tau_2}\right)^{\frac{r}{2}}
\sum_{w, m\in Q^\vee}
\xi_{m+\mu,w+\omega} H_{m+\mu,w+\omega},
\end{align}
with 
\begin{align}
\xi_{m,w} & = e^{-\pi i\kappa(m,w)}, \label{eq:xiPW}\\
H_{m,w} & = 
\exp\{
-\frac{\pi\kappa}{2\tau_2}
\big(
(m-\tau w+2u,\,
m-\bar{\tau}w-2\bar v) \notag \\
& \hspace{4cm }+2 (u, \bar v) + (u, \bar u) + (\bar v, v)
\big)
\}. \label{eq:HkinGauging}
\end{align}
Combining this result with the orbifold sum and Weyl-group sum gives the localization formula for the SWZW partition function,
\begin{align}
Z_{\mathrm{SWZW}}
&=
2^r\,
{\rm Vol}(\qqv)
\left(
\frac{\kappa}{2\tau_2}
\right)^{\frac{r}{2}}
\sum_{\sigma\in W}
\det(\sigma)
\frac{1}{\lvert\mathcal{C}\rvert}
\sum_{\mu,\omega\in\mathcal{C}}
\Phi_{\mu,\omega}
\sum_{m,w \in \qqv} \xi_{m+\mu,w+\omega}\,
H_{m +\mu,w+\omega}^\sigma.
\end{align}
Finally, combining the sums over $Q^\vee$ with the sector labels
$\mu,\omega\in\mathcal C\simeq\Lambda^\vee/Q^\vee$ promotes the
coroot-lattice sums to sums over the full cocharacter lattice
$\Lambda^\vee$. We obtain
\begin{align}
\label{eq:localizationHamilCocha}
 Z_{\rm SWZW}=
2^r\,
{\rm Vol}(\Lambda^\vee)
\left(
\frac{\kappa}{2\tau_2}
\right)^{\frac{r}{2}}
\sum_{\sigma\in W}
\det(\sigma)
\sum_{m,w \in \Lambda^\vee}
\Phi_{[m],[w]}\, \xi_{m,w}\,
H_{m,w}^\sigma.
\end{align}
where
$[m],[w]\in\Lambda^\vee/Q^\vee\simeq\mathcal C$.

This is the non-simply connected Murthy--Witten localization formula,
supplemented by the phase $\Phi_{[m],[w]}$ inherited
from the FGK modular invariant. As a consistency check, consider the simply connected case. Then
$\mathcal{C}$ is trivial and $
\Lambda^\vee=\qqv$, so the localization formula reduces to
\begin{align}
Z_{\mathrm{SWZW}}(\tau, u, v)
=
2^r\,
\text{Vol}( \qqv)
\left(
\frac{\kappa}{2\tau_2}
\right)^{\frac{r}{2}}
\sum_{\sigma\in W}
\det(\sigma)
\sum_{m,w\in \qqv} \xi_{m,w}
H_{m,w}^\sigma,
\end{align}
which precisely reproduces the result in~\cite{Lu_wzwlocalization2026}.

\subsection{New phases}
\label{subsec:newphases}

We now examine the structure of the phase
$\Phi_{\mu,\omega}$ defined in~\eqref{eq:Phi}.
For $G_2,F_4,E_8$ the center is trivial and hence $\Phi=1$.
Explicit expressions for all global forms are collected in
Appendix~\ref{apx:phasePhi}.

For a cyclic quotient $\mathcal C=\mathbb Z_N$ generated by the
minuscule coweight class $\gamma$, writing
\begin{align}
\mu=m\gamma,
\quad
\omega=n\gamma,
\quad
m,n\in\mathbb Z_N,
\end{align}
gives the universal expression
\begin{align}
\label{eq:Phi-cyclic}
\Phi^{\mathfrak g,\,\mathcal C}_{m,n}
=
(-1)^{h^\vee\gamma^2(m^2+n^2)}
\exp\{
\pi i k\gamma^2mn
\}.
\end{align}
The only noncyclic case is the full center
$\mathcal C=\mathbb Z_2\times\mathbb Z_2$ of $D_{2r}$, for which
the phase contains in addition a discrete torsion factor. We refer to Appendix~\ref{apx:phasePhi}
for its explicit form.

When $k$ obeys the CS constraint, the $k$-dependent bilinear phase,
together with the discrete torsion factor in the $D_{2r}$ full-center
case, is trivial. The surviving phase is therefore the
$h^\vee$-dependent diagonal sign. Since $h^\vee \mu^2 = (2\rho,\mu)$, and 
$e^{2\pi i(\rho,\gamma)}=1$ for $\gamma\in Q^\vee$, this sign admits
a natural lift from $\mathcal C=\Lambda^\vee/Q^\vee$ to the
cocharacter lattice $\Lambda^\vee$, which we denote by
\begin{align}
\Phi_{m,w}
=
e^{2\pi i (\rho,m +w)},
\quad
m,w\in\Lambda^\vee.
\end{align}
Consequently, the SWZW partition function becomes
\begin{align}
\label{eq:partitionCS}
 Z_{\rm SWZW}=
2^r\,
{\rm Vol}(\Lambda^\vee)
\left(
\frac{\kappa}{2\tau_2}
\right)^{\frac{r}{2}}
\sum_{\sigma\in W}
\det(\sigma)
\sum_{m,w \in \Lambda^\vee}
\Phi_{m,w}\, \xi_{m,w}\,
H_{m,w}^\sigma.
\end{align}

\section{Localization and abelianization}
\label{sec:localization}

In this section we extend the localization computation of the SWZW model in~\cite{murthyW25, Lu_wzwlocalization2026} to compact connected simple Lie groups that are not necessarily simply connected. The localization argument and the abelianization of the localizing solutions remain unchanged, while additional topological sectors associated with non-trivial fundamental groups arise. Further effects of the nontrivial topology enter through the classical WZ amplitude and the fermion one-loop determinant. We will show that the former acquires the FGK cocycle phase, while the latter acquires a global-anomaly phase captured by the relative Rochlin invariants. Together, these phases reproduce the additional phase factors identified in the previous section.

\subsection{Gauged SWZW and supersymmetric localization}

The SWZW model on $T^2$ consists of a group-valued field $g$ and
adjoint-valued Majorana--Weyl fermions $\psi$ and $\tilde\psi$. Its action
decomposes into bosonic and fermionic parts as
\begin{align}
I_{\rm SWZW}(g,\psi,\tilde\psi)
=
I_{\rm WZW}(g)+I_F(\psi)+I_F(\tilde\psi).
\end{align}
The bosonic WZW action is
\begin{align}
I_{\rm WZW}(g)
=
I_{\rm kin}(g)+i\Gamma_{\rm WZ}(g),
\end{align}
where the sigma-model kinetic term is
\begin{align}
I_{\rm kin}(g)
&=
-\frac{i}{4\pi}
\int_{T^2}dz\wedge d\bar z\,
\tr\left(
g^{-1}\partial_z g\,
g^{-1}\partial_{\bar z}g
\right),
\end{align}
and the WZ term is
\begin{align}
\Gamma_{\rm WZ}(g)
&=
\frac{1}{12\pi}
\int_B
\tr(g^{-1}dg)^3,
\qquad
\partial B=T^2.
\end{align}
The two Majorana--Weyl fermions have opposite chiralities, with actions
\begin{align}
I_F(\psi)
&=
-\frac{1}{4\pi}
\int_{T^2}dz\wedge d\bar z\,
\tr \psi\partial_{\bar z}\psi,
\\
I_F(\tilde\psi)
&=
-\frac{1}{4\pi}
\int_{T^2}dz\wedge d\bar z\,
\tr \tilde\psi\partial_z\tilde\psi .
\end{align}
The SWZW action is invariant under the global $G^L\times G^R$ symmetry
\begin{align}
g &\mapsto h_L g h_R^{-1},
\qquad
\tilde\psi \mapsto h_L\tilde\psi h_L^{-1},
\qquad
\psi \mapsto h_R\psi h_R^{-1}.
\end{align}
To couple this symmetry to background gauge fields $A^L$ and $A^R$, we
introduce the covariant derivatives
\begin{align}
Dg &= dg+A^L g-gA^R,
\qquad
D\tilde\psi = d\tilde\psi+[A^L,\tilde\psi],
\qquad
D\psi = d\psi+[A^R,\psi].
\end{align}
We take the background gauge fields to be flat and Cartan-valued.
Writing $z=s+\tau t$, with $s,t$ of period $2\pi$ and
$ds^2=|dz|^2=|ds+\tau dt|^2$, we parametrize them as
\begin{align}
\label{eq:ALAR}
A^L
&=
\frac{\bar v\,dz-v\,d\bar z}{\tau-\bar\tau},
\qquad
A^R
=
\frac{\bar u\,dz-u\,d\bar z}{\tau-\bar\tau}.
\end{align}
In these backgrounds, the gauged SWZW action takes the form
\begin{align}
I_{\rm SWZW}(g,\psi,\tilde\psi;\,A^L,A^R)
=
I_{\rm WZW}(g,A^L,A^R)
+I_F(\psi,A^R)
+I_F(\tilde\psi,A^L),
\end{align}
where the gauged WZW action is
\begin{align}
I_{\rm WZW}(g,A^L,A^R)
=
I_{\rm kin}^A(g,A^L,A^R)
+i\Gamma_{\rm WZ}(g).
\end{align}
The kinetic term and background gauge couplings are collected in
\begin{align}
I_{\rm kin}^A(g,A^L,A^R)
&=
I_{\rm kin}(g)
-\frac{i}{2\pi}
\int_{T^2}dz\wedge d\bar z\,
\tr\Big[
A_z^L\partial_{\bar z}g\,g^{-1}
-g^{-1}\partial_zg\,A_{\bar z}^R
-A_z^LgA_{\bar z}^Rg^{-1}
\nonumber\\
&\hspace{8.0cm}
+\frac12
\left(
A_z^LA_{\bar z}^L
+A_z^RA_{\bar z}^R
\right)
\Big],
\end{align}
while the gauged fermion actions are
\begin{align}
I_F(\psi,A^R)
&=
-\frac{1}{4\pi}
\int_{T^2}dz\wedge d\bar z\,
\tr \psi D_{\bar z}\psi ,
\\
I_F(\tilde\psi,A^L)
&=
-\frac{1}{4\pi}
\int_{T^2}dz\wedge d\bar z\,
\tr \tilde\psi D_z\tilde\psi .
\end{align}
For localization, we choose the supersymmetry transformations
\begin{align}
Qg &= ig\psi,
\qquad
Q\psi = \mathcal J_z-i\psi^2,
\qquad
Q\tilde\psi = 0,
\qquad
QA^L=QA^R=0,
\end{align}
where $\mathcal J_z=g^{-1}D_zg$. The supersymmetry algebra closes off shell as
\begin{align}
Q^2=iD_z.
\end{align}

We consider the torus partition function of the SWZW model with periodic
boundary conditions for the fermions. The resulting Cartan-valued fermion
zero modes are saturated by the insertion
\begin{align}
\Xi =
(\psi_0^1\cdots\psi_0^r)
(\tilde\psi_0^1\cdots\tilde\psi_0^r),
\end{align}
where $\psi_0^i$ and $\tilde\psi_0^i$ denote the constant zero modes of the
Cartan components of $\psi(z)$ and $\tilde\psi(\bar z)$, respectively.
Following~\cite{murthyW25}, we choose the real Clifford-module convention
in which each left-right Majorana zero-mode pair has grading operator
\begin{align}
(-1)^{F_i}
=
i\psi_0^i\tilde\psi_0^i,
\qquad
\big((-1)^{F_i}\big)^2=1,
\end{align}
with total fermion parity
\begin{align}
(-1)^F
=
\prod_{i=1}^r(-1)^{F_i}.
\end{align}
With this choice of zero-mode normalization, the partition function is
\begin{align}
Z_{\rm SWZW}
=
\int
\mathcal Dg\,\mathcal D\psi\,\mathcal D\tilde\psi\;
\Xi\,
\exp\{-\kappa I_{\rm SWZW}\}.
\end{align}
The fermions induce the standard one-loop shift of the level,
$\kappa=k+h^\vee$. The bare WZW level $k$ must satisfy the WZW quantization constraints summarized in Appendix~\ref{apx:constraints}. Since $h^\vee$ satisfies the same constraints,
the shifted level $\kappa$ obeys the same global quantization conditions.

The fermion zero-mode insertion constrains the choice of localizing
deformation. A convenient choice is
\begin{align}
V
=
i \int_{T^2}dz\wedge d\bar z\,
\tr\left(
(D_{\bar z}\psi)(D_z\mathcal J_{\bar z})
\right),
\end{align}
for which the $Q$-exact deformation is
\begin{align}
QV
&=
i\int_{T^2}dz\wedge d\bar z\,
\tr\left(
D_{\bar z}(\mathcal J_z-i\psi^2)
D_z\mathcal J_{\bar z}
\right)
\nonumber\\
&\hspace{4cm}
-
\int_{T^2}dz\wedge d\bar z\,
\tr\left(
D_{\bar z}\psi\,
D_z\left(D_{\bar z}\psi+[\mathcal J_{\bar z},\psi]\right)
\right).
\end{align}
Its bosonic part is
\begin{align}
(QV)_{\rm bos}
=
i\int_{T^2}dz\wedge d\bar z\,
\tr\left(
D_{\bar z}\mathcal J_z
D_z\mathcal J_{\bar z}
\right)
\geq0 .
\end{align}
We therefore consider the deformed partition function
\begin{align}
Z_V(\lambda)
=
\int
\mathcal Dg\,\mathcal D\psi\,\mathcal D\tilde\psi\;
\Xi\,
\exp\{
-\kappa I_{\rm SWZW}-\lambda QV
\}.
\end{align}
In the limit $\lambda\to\infty$, the path integral localizes onto the
bosonic configurations satisfying
\begin{align}
D_{\bar z}\mathcal J_z=0.
\end{align}
This equation states that $g_{\rm cl}$ defines a holomorphic
isomorphism between the two complexified adjoint bundles~\cite{murthyW25, Lu_wzwlocalization2026}. As in the
simply connected case, the localization locus can be abelianized,
so that
\begin{align}
g_{\rm cl}=g_\sigma\hat g,
\quad
\sigma\in W,
\end{align}
with $\hat g$ valued in the maximal torus $T$.
On a torus $T^2$,  the maximal torus valued factor can be written as
\begin{align}
    \tilde g =\exp\{i (m t +ws)\}.
\end{align}
Periodicity of $\tilde g$ implies that $m$, $w$ lie in the cocharacter lattice. 
Hence, the localization solutions are labeled by $(\sigma,m,w)\in W\times \Lambda^\vee\times \Lambda^\vee$. 
Explicitly,
\begin{align}
g_{\rm cl}
=
g_\sigma e^{i(mt+ws)}
=
g_\sigma
\exp\{
i\frac{\bar\alpha z-\alpha\bar z}
{\tau-\bar\tau}
\},
\quad
\alpha=m-w\tau.
\end{align}
We next expand around a localizing solution as
\begin{align}
g=g_{\rm cl}e^{y/\sqrt{\lambda}},
\end{align}
with the fermion simultaneously rescaled as
$\psi\to\psi/\sqrt{\lambda}$.
Then
\begin{align}
\sqrt{\lambda}\,\mathcal J_z
=
\sqrt{\lambda}\,\mathcal J_z^{\rm cl}
+
\left(
\partial_z
+ (A^L_z)_{g_{\rm cl}}
\right)y
+O(\lambda^{-1/2}),
\end{align}
where
\begin{align}
    \mathcal J_z^{\rm cl}= (A^L_z)_{g_{\rm cl}}-A_z^R, 
    \quad (A^L_z)_g=g^{-1}(\pd_z+A^L_z)g.
\end{align}
Since $D_{\bar z}\mathcal J_z^{\rm cl}=0$,
\begin{align}
\sqrt{\lambda}\,
D_{\bar z}\mathcal J_z
=
D_{\bar z}
\left(
\partial_z + (A^L_z)_{g_{\rm cl}} 
\right)y
+O(\lambda^{-1/2}).
\end{align}
Consequently, at quadratic order,
\begin{align}
\kappa I_{\rm SWZW}+\lambda QV
=
\kappa I_{\rm WZW}(g_{\rm cl},A^L,A^R)
+\kappa I_F(\tilde\psi,A^L)
+(QV)^{(2)}
+O(\lambda^{-1/2}).
\end{align}
Taking $\lambda\to\infty$, the partition function therefore reduces to
a sum over the localization locus,
\begin{align}
Z_{\rm SWZW}
=
\sum_{\sigma\in W}
\sum_{m,w\in\Lambda^\vee}
Z^{\rm cl}_{\sigma,m,w}\,
Z^{\text{1-loop}}_{\sigma,m,w},
\end{align}
with
\begin{align}
Z^{\rm cl}_{\sigma,m,w}
&=
e^{-\kappa I_{\rm WZW}(g_{\rm cl},A^L,A^R)} =
A_{\rm WZ}(g_{\rm cl})\exp\{-\kappa I_{\rm kin}^A(g_{\rm cl},A^L,A^R)\},\\
Z^{\text{1-loop}}_{\sigma,m,w}
&=
\int\mathcal D\tilde\psi\,
e^{-\kappa I_F(\tilde\psi,A^L)}
\int\mathcal Dy\,\mathcal D\psi\,
e^{-(QV)^{(2)}} .
\end{align}

\subsection{WZ amplitude and FGK cocycle}

The WZ amplitude on the localizing solutions can be evaluated using the
generalized PW formula~\cite{FGK1988}. For maps $g_1$ and $g_2$ in homotopy classes
$[\mu,\omega], [\mu',\omega'] \in \mathcal{C} \times \mathcal{C}$, respectively, the WZ amplitude obeys
\begin{align}
\label{eq:generalizedPW}
A_{\rm WZ}[g_1g_2]
&=
c^{[\mu,\omega],[\mu',\omega']}_{\rm WZ}
{\rm \Phi}_{\rm WZ}(g_1, g_2)
A_{\rm WZ}[g_1]A_{\rm WZ}[g_2],
\end{align}
where the FGK cocycle is
\begin{align}
\label{eq:FGKcocycle}
c^{[\mu,\omega], [\mu',\omega']}_{\rm WZ}
=
(-1)^{\epsilon(\mu\wedge\omega'+\mu'\wedge\omega)}
e^{
\pi i\kappa
\left(
(\mu,\omega')-(\omega,\mu')
\right)
},
\end{align}
and the PW cocycle is 
\begin{align}
\label{eq:PWcocycle}
   {\rm \Phi}_{\rm WZ}(g_1,g_2) =  \exp\{
\frac{i\kappa}{4\pi}
\int_{T^2}
\tr\left(
g_1^{-1}dg_1\wedge dg_2g_2^{-1}
\right)\}.
\end{align}
For the localizing solution $g_{\rm cl}=g_\sigma g_1g_2$, we take $g_1=e^{imt}$ and $g_2=e^{iws}$. The two factors represent the homotopy classes $[[m],0]$ and $[0,[w]]$, respectively.
The constant factor $g_\sigma$ does not contribute to the WZ amplitude, hence
\begin{align}
    A_{\rm WZ}(g_{\rm cl})=A_{\rm WZ}(g_1 g_2).
\end{align}
Moreover, since $g_1$ and $g_2$ each depend on only one torus coordinate,
\begin{align}
A_{\rm WZ}(g_1)=A_{\rm WZ}(g_2)=1.
\end{align}
The two cocycle factors evaluate to
\begin{align}
{\rm \Phi}_{\rm WZ}(g_1,g_2)=\xi_{m,w},
\quad
c^{[[m],0 ], \,[0,[w]]}_{\rm WZ}=\phi_{[m],[w]},
\end{align}
where $\xi_{m,w}$ is given by~\eqref{eq:xiPW}, and
\begin{align}
\label{eq:phiphase}
   \phi_{[m],[w]} = (-1)^{\epsilon[m]\wedge[w]}
e^{\pi i\kappa([m],[w])}.
\end{align}
The FGK factor $\phi_{[m],[w]}$ is precisely the KSB~\eqref{eq:KSB}, evaluated at the shifted level $\kappa$.
Thus the WZ amplitude factorizes into the FGK contribution $\phi_{[m],[w]}$, which depends only on the homotopy classes, and the PW contribution $\xi_{m,w}$,
\begin{align}
     A_{\rm WZ}(g_{\rm cl}) = \phi_{[m],[w]} \xi_{m,w}.
\end{align}
The kinetic and gauge-coupling part of the gauged WZW action evaluates to
\begin{align}
\exp\{
-\kappa I_{\rm kin}^A(g_{\rm cl},A^L,A^R)
\}
=
H_{m,w}^\sigma.
\end{align}
Here $H_{m,w}$ is given by~\eqref{eq:HkinGauging}, with $H^\sigma_{m,w}(\tau, u, v) = H_{m,w}(\tau, \sigma(u), v)$.
Combining this with the WZ amplitude, the classical contribution is therefore
\begin{align}
Z^{\rm cl}_{\sigma,m,w}
= \phi_{[m],[w]}\,
\xi_{m,w}\,
H_{m,w}^\sigma.
\end{align}

\subsection{Pfaffian phase and global anomaly}
\label{subsec:pfaffian}
We now determine the one-loop contribution. We first separate its
dependence on the Weyl-group element $\sigma$ from that on the winding
data $m$ and $w$. The nonzero-mode fluctuation measure is invariant under
the Weyl action, while the Cartan fermion zero-mode insertion transforms
by the determinant of the Weyl action on $\mathfrak h$. The Weyl-group dependence therefore
factorizes as
\begin{align}
Z^{\text{1-loop}}_{\sigma,m,w}
=
\det(\sigma)\,
Z^{\text{1-loop}}_{m,w}.
\end{align}

The remaining contribution is invariant under coroot shifts
$m\to m+\beta$ and $w \to w+\gamma$, with $\beta,\gamma\in Q^\vee$.
Locally, such shifts merely relabel the root-space fluctuation modes,
while globally they lift to closed transformations in the simply
connected cover $\widetilde G$ and hence have trivial relative
Pfaffian holonomy. Consequently, the one-loop contribution depends on
$m$ and $w$ only through the classes
$[m],[w]\in
\Lambda^\vee/Q^\vee
\simeq\mathcal C$, and therefore
\begin{align}
Z^{\text{1-loop}}_{\sigma,m,w}
=
\det(\sigma)\,
Z^{\text{1-loop}}_{[m],[w]}.
\end{align}

In the trivial winding class, the one-loop contribution is the same
as in the simply connected theory, apart from the volume of the
maximal torus appropriate to the chosen global form,
\begin{align}
Z^{\text{1-loop}}_{0,0}
=
2^r\,
{\rm Vol}(\Lambda^\vee)
\left(
\frac{\kappa}{2\tau_2}
\right)^{r/2}.
\end{align}
It is therefore convenient to write
\begin{align}
Z^{\text{1-loop}}_{\sigma,m,w}
=
2^r\,\det(\sigma)
{\rm Vol}(\Lambda^\vee)
\left(
\frac{\kappa}{2\tau_2}
\right)^{r/2}
\nu_{[m],[w]},
\end{align}
where
\begin{align}
\nu_{[m],[w]}
=
\frac{
Z^{\text{1-loop}}_{[m],[w]}
}{
Z^{\text{1-loop}}_{[0],[0]}
}.
\end{align}
Here the ratio is understood as the relative one-loop holonomy between
the winding sector $([m],[w])$ and the trivial sector. The
winding-dependent part of the one-loop contribution arises from the
Pfaffian of the Majorana--Weyl fermion $\psi$, so that
$\nu_{[m],[w]}$ is equivalently the corresponding relative Pfaffian
holonomy.
Indeed, for Majorana--Weyl fermions the Pfaffian is not in general a
single-valued function on the space of background gauge fields, but is
naturally a section of a Pfaffian line bundle equipped with the
Bismut--Freed connection~\cite{BismutFreed1986I}. The Pfaffian line
bundles associated with different winding sectors have the same local
curvature, and hence their relative Pfaffian line bundle is flat.
Consequently, the local anomaly cancels in the relative Pfaffian, while
a global anomaly may remain as the holonomy of this flat line
bundle~\cite{BismutFreed1986II,LeeMillerWeintraub1988}.

Thus the winding-dependent factor $\nu_{[m],[w]}$ is the relative
Pfaffian holonomy between the sector labeled by $[m],[w]\in\mathcal C$
and the trivial sector. Since the relative Pfaffian line bundle is flat,
this holonomy depends only on the homotopy class of the corresponding
loop in background-field space. By the Bismut--Freed holonomy
theorem~\cite{BismutFreed1986II}, it may be expressed in terms of the reduced eta invariant of
Atiyah--Patodi--Singer of the Dirac operator on the corresponding
mapping torus~\cite{APS1975Ireducedeta},
\begin{align}
\nu_{[m],[w]}
=
\exp\{
- \pi i
\left(
\bar\eta(D_{[m],[w]})
-
\bar\eta(D_{[0],[0]})
\right)
\}.
\end{align}
Here
\begin{align}
\label{eq:reducedEta}
\bar\eta(D)
=
\frac{1}{2}
\left(
\eta(D)+\dim\ker D
\right)
\end{align}
denotes the reduced eta invariant.
In the Dai--Freed formulation~\cite{DaiFreed1994eta,Witten2015FermionPI,
Yonekura2016DaiFreed,WittenYonekura2019eta}, this holonomy represents
the global anomaly of the fermion Pfaffian line bundle.

To evaluate this holonomy explicitly, we decompose the non-Cartan part
of the adjoint representation into root spaces. The relative Pfaffian
holonomy factorizes accordingly as
\begin{align}
\nu_{[m],[w]}
=
\prod_{\alpha\in R}
\nu^\alpha_{[m],[w]}.
\end{align}
For each root $\alpha$, choose coweight representatives $\mu,\omega$
of the winding classes $[m],[w]\in\mathcal C$. Their mod-$2$ pairings
with $\alpha$ determine the twist
\begin{align}
s^\alpha
=
s_a^\alpha a+s_b^\alpha b
\in H_1(T^2,\mathbb Z_2),
\end{align}
where
\begin{align}
s_a^\alpha
=
(\alpha,\mu)\pmod 2,
\qquad
s_b^\alpha
=
(\alpha,\omega)\pmod 2.
\end{align}

The untwisted fermion has PP spin structure, while $s^\alpha$ shifts
the spin structure relative to PP. Denoting the corresponding quadratic
refinements by $q^{\rm PP}$ and $q^{\rm PP}_{s^\alpha}$, respectively,
the root-space contribution to the relative Pfaffian holonomy is
\begin{align}
\nu^\alpha_{[m],[w]}
:=
\frac{
{\rm Pf}(q^{\rm PP}_{s^\alpha})
}{
{\rm Pf}(q^{\rm PP})
}.
\end{align}
Here the ratio denotes the relative Pfaffian holonomy between the two
spin structures, with the Pfaffian fibers identified by parallel
transport.

The change of the Arf invariant under the twist $s^\alpha$ is determined
by the quadratic refinement,
\begin{align}
q^{\rm PP}(s^\alpha)
=
(\alpha,\mu)
+
(\alpha,\omega)
+
(\alpha,\mu)(\alpha,\omega)
\pmod 2.
\end{align}
As reviewed in Appendix~\ref{apx:arfRochlin}, the relative Rochlin
invariant determines this Pfaffian holonomy, and for the PP spin
structure one finds~\cite{LeeMillerWeintraub1988}
\begin{align}
\nu^\alpha_{[m],[w]}
=
\frac{
{\rm Pf}(q^{\rm PP}_{s^\alpha})
}{
{\rm Pf}(q^{\rm PP})
}
=
(-i)^{q^{\rm PP}(s^\alpha)}.
\end{align}
The root spaces $\mathfrak g_\alpha$ and $\mathfrak g_{-\alpha}$ are
paired by the Majorana reality condition and together form a real
two-plane. Since $s^{-\alpha}=s^\alpha$, one has
$q^{\rm PP}(s^{-\alpha})=q^{\rm PP}(s^\alpha)$, and hence
\begin{align}
\nu^{-\alpha}_{[m],[w]}
\nu^\alpha_{[m],[w]}
&=
(-1)^{q^{\rm PP}(s^\alpha)}
\nonumber\\
&=
(-1)^{
(\alpha,\mu)(\alpha,\omega)
+
(\alpha,\mu)
+
(\alpha,\omega)
}.
\end{align}
Thus each real root plane
$\mathfrak g_\alpha\oplus\mathfrak g_{-\alpha}$ contributes a real
sign. Multiplying over the positive roots gives
\begin{align}
\nu_{[m],[w]}
=
(-1)^{
\sum_{\alpha>0}
\left[
(\alpha,\mu)(\alpha,\omega)
+
(\alpha,\mu)
+
(\alpha,\omega)
\right]
}.
\end{align}
Using
\begin{align}
\sum_{\alpha>0}
(\alpha,\mu)(\alpha,\omega)
&=
h^\vee(\mu,\omega),
\nonumber\\
(2\rho,\mu)
&=
h^\vee\mu^2,
\qquad
(2\rho,\omega)
=
h^\vee\omega^2,
\end{align}
we obtain the global fermionic phase
\begin{align}
\label{eq:fermion-global-phase}
\nu_{[m],[w]}
=
(-1)^{
h^\vee\left[
(\mu,\omega)+\mu^2+\omega^2
\right]
}.
\end{align}
The sign $\nu_{[m],[w]}$ is therefore the product of the relative
Pfaffian holonomies of the real root planes. Since the underlying
quadratic-refinement data are mod-$2$, this global fermionic phase
detects only the $2$-primary part of $\mathcal C$ and is therefore
trivial for quotients of odd order. The case-by-case triviality
conditions are collected in Appendix~\ref{apx:phaseNu}.

\subsection{The localization formula}
Combining the classical and one-loop contributions, we obtain the SWZW
partition function
\begin{align}
Z_{\rm SWZW}
&=
\sum_{\sigma\in W}
\sum_{m,w\in\Lambda^\vee}
Z^{\rm cl}_{\sigma,m,w}
Z^{\rm 1-loop}_{\sigma,m,w}
\nonumber\\
&=
2^r\, 
{\rm Vol}(\Lambda^\vee)
\left(
\frac{\kappa}{2\tau_2}
\right)^{r/2}
\sum_{\sigma\in W}\det(\sigma)
\sum_{m,w\in\Lambda^\vee}
\phi_{[m],[w]}\,
\nu_{[m],[w]}\,
\xi_{m,w}
H_{m,w}^\sigma .
\end{align}
Here $\phi_{[m],[w]}$ is the FGK/KSB phase~\eqref{eq:phiphase},
$\nu_{[m],[w]}$ is the fermionic Pfaffian phase~\eqref{eq:fermion-global-phase},
and $\xi_{m,w}$ is the PW cocycle phase~\eqref{eq:xiPW}.
Using $\kappa=k+h^\vee$ and $h^\vee([m],[w])\in\bz$, we have
\begin{align}
e^{\pi i\kappa([m],[w])}
=
e^{\pi i k([m],[w])}
(-1)^{h^\vee([m],[w])}.
\end{align}
It follows that $\phi$ and $\nu$ combine to give $\Phi$ in~\eqref{eq:Phi},
\begin{align}
\phi_{[m],[w]}\nu_{[m],[w]}
&=
(-1)^{\epsilon[m]\wedge[w]}
(-1)^{h^\vee([m]^2+[w]^2)}
e^{\pi i k([m],[w])}
\nonumber\\
&=
\Phi_{[m],[w]}.
\end{align}
Hence
\begin{align}
Z_{\rm SWZW}
=
2^r\,{\rm Vol}(\Lambda^\vee)
\left(
\frac{\kappa}{2\tau_2}
\right)^{r/2}
\sum_{\sigma\in W}
\det(\sigma)
\sum_{m,w\in\Lambda^\vee}
\Phi_{[m],[w]}\,
\xi_{m,w}\,
H_{m,w}^\sigma,
\end{align}
which reproduces the localization
formula~\eqref{eq:localizationHamilCocha} derived from the Hamiltonian
formulation. Thus, the additional phase $\Phi_{[m],[w]}$ originally
found in the Hamiltonian formalism has a direct path-integral origin:
it arises from the combination of the topological phase of the classical
action and the global Pfaffian anomaly of the chiral fermions.

\section{Narain lattice and simple-current orbifolds}
\label{sec:narain}

In this section we give an abelian interpretation of the topological phases appearing in the non-simply connected SWZW model. We first generalize the $B$-field amplitude on a toroidal target space to sectors with nontrivial shift holonomies and derive the corresponding abelian PW formula. We then show that the restriction of the WZ amplitude to the maximal torus takes precisely this form: the continuous part is described by a constant Kalb–Ramond (KR) $B$-field, while the homotopy-dependent FGK cocycle becomes the flat-gerbe data of the abelian theory. Finally, we rewrite the SWZW partition function as a sum of Siegel–Narain theta functions associated with shifted Narain lattices.

\subsection{Generalized $B$-field amplitudes}
\label{subsec:generalizedBamp}

Consider the Narain torus $T^d=\mathbb R^d/(2 \pi L)$, equipped with metric
$G_{ij}$ and a constant KR $B$-field $B_{ij}$. We refer to \cite{Lu_wzwlocalization2026} for a review of the Narain torus
theory and our conventions. We choose an intermediate lattice
\begin{align}
L\subset M\subset L^\ast,
\quad
S=M/L,
\end{align}
so that the finite abelian group $S$ acts on $T^d$ by shifts. Gauging
this symmetry decomposes the theory into twisted sectors labeled by
$\mu,\omega\in S$, with boundary conditions on a torus $T^2$
\begin{align}
X(t+2\pi,s)
&=
X(t,s)+2\pi (m+\mu),
&
X(t,s+2\pi)
&=
X(t,s)+2 \pi (w+\omega),
\quad m,w\in L.
\end{align}

We first recall the ordinary $B$-field holonomy. For an untwisted map
$X:T^2 \to T^d$, it is given by
\begin{align}
A_{\rm KR}(X)
=
\exp\{
-\frac{i}{8\pi}
\int_\Sigma
B_{ij}\,dX^i\wedge dX^j
\},
\end{align}
and satisfies the abelian Polyakov--Wiegmann identity~\cite{Lu_wzwlocalization2026}
\begin{align}
A_{\rm KR}(X+X')
=
\Phi_{\rm KR}(X,X')
A_{\rm KR}(X)A_{\rm KR}(X'),
\end{align}
where
\begin{align}
  \Phi_{\rm KR}(X,X') =  \exp\{
-\frac{i}{4\pi}
\int_\Sigma
B_{ij}\,dX^i\wedge dX'^j
\}.
\end{align}
In the twisted sectors, extending the $B$-field amplitude to torsion
cycles introduces an additional topological ambiguity corresponding
to a flat gerbe. On the torsion sectors this ambiguity is specified by
a character
\begin{align}
\chi:H_2(S,\mathbb Z)\longrightarrow U(1).
\end{align}
For the WZW applications below it suffices to consider
$S=\mathbb Z_N$ and $S=\mathbb Z_N\times\mathbb Z_N$. For
$S=\mathbb Z_N$,
\begin{align}
H_2(\mathbb Z_N,\mathbb Z)=0,
\end{align}
and hence there is no additional torsion ambiguity. For
$S=\mathbb Z_N\times\mathbb Z_N$, instead,
\begin{align}
H_2(S,\mathbb Z)\simeq\mathbb Z_N.
\end{align}

Let $e_1\wedge e_2$ denote the generator of $H_2(S,\mathbb Z)$ and
consider the corresponding torsional map
\begin{align}
X_0=t e_1+s e_2.
\end{align}
Since the homology class of $X_0$ has order $N$, consistency with the
ordinary $B$-field holonomy fixes the $N$-th power of its amplitude,
\begin{align}
A_{\rm KR}(X_0)^N
=
e^{-i\pi N B(e_1,e_2)}.
\end{align}
The possible extensions to $X_0$ therefore differ by an $N$-th root
of unity,
\begin{align}
A_{\rm KR}(X_0)
=
\zeta\,e^{-i\pi B(e_1,e_2)},
\quad
\zeta^N=1.
\end{align}
The ambiguity $\zeta$ is the value of the flat-gerbe character on the
generator,
\begin{align}
\zeta=\chi(e_1\wedge e_2).
\end{align}

It is convenient to denote the full torsional amplitude of the
generator by
\begin{align}
a_B
\equiv
A_{\rm KR}(X_0)
=
\chi(e_1\wedge e_2)\,
e^{-i\pi B(e_1,e_2)}.
\end{align}
For general torsional representatives,
\begin{align}
A_{\rm KR}(X_{\mu,\omega})
=
a_B^{\,\mu\wedge\omega}.
\end{align}
Changing the lift by $N$ changes the amplitude by the corresponding
ordinary $B$-field holonomy, consistently with
\begin{align}
a_B^N
=
e^{-i\pi N B(e_1,e_2)}.
\end{align}

Using the bilinearity of the wedge product, we then find
\begin{align}
\frac{
A_{\rm KR}(X_{\mu,\omega}+X_{\mu',\omega'})
}{
A_{\rm KR}(X_{\mu,\omega})
A_{\rm KR}(X_{\mu',\omega'})
}
=
a_B^{\,\mu\wedge\omega'+\mu'\wedge\omega}.
\end{align}
Combining this torsional contribution with the ordinary $B$-field
Polyakov--Wiegmann factor gives, for arbitrary maps $X$ and $X'$ with
torsion classes $(\mu,\omega)$ and $(\mu',\omega')$,
\begin{align}
A_{\rm KR}(X+X')
&=
a_B^{\,\mu\wedge\omega'+\mu'\wedge\omega}
\exp\{
\frac{i}{4\pi}
\int_\Sigma
B_{ij}\,
dX_{\mu,\omega}^i\wedge dX_{\mu',\omega'}^j
\}
\nonumber\\
&\quad\times
\Phi_{\rm KR}(X,X')
A_{\rm KR}(X)A_{\rm KR}(X').
\end{align}
For the harmonic representatives on $T^2$,
\begin{align}
\exp\{
\frac{i}{4\pi}
\int_{T^2}
B_{ij}\,
dX_{\mu,\omega}^i\wedge dX_{\mu',\omega'}^j
\}
=
\exp\{
i\pi\left(
B(\mu,\omega')-B(\omega,\mu')
\right)
\}.
\end{align}
With
\begin{align}
\label{eq:abelianFGKphaseKR}
    c_{\rm KR}^{[\mu,\omega],[\mu',\omega']} = a_B^{\,\mu\wedge\omega'+\mu'\wedge\omega} \exp\{
i\pi\left(
B(\mu,\omega')-B(\omega,\mu')
\right)
\},
\end{align}
we therefore obtain the generalized abelian Polyakov--Wiegmann formula
\begin{align}
A_{\rm KR}(X+X')
=c_{\rm KR}^{[\mu,\omega],[\mu',\omega']} 
\Phi_{\rm KR}(X,X')
A_{\rm KR}(X)A_{\rm KR}(X').
\label{eq:generalized-abelian-PW}
\end{align}
For the WZW application below, we further restrict the $B$-field data by a
reality condition. Since under $X\mapsto-X$ the generator $X_0$ is related to
itself by the orientation-preserving reparameterization
$(t,s)\mapsto(-t,-s)$, reality requires
\begin{align}
A_{\rm KR}(X_0)=\overline{A_{\rm KR}(X_0)}.
\end{align}
We therefore restrict to $B$-field data for which
\begin{align}
a_B
=
\chi(e_1\wedge e_2)e^{-i\pi B(e_1,e_2)}
=
(-1)^\epsilon, \quad \epsilon \in \{0,1\}.
\end{align}
For odd $N$ the nontrivial sign is incompatible with the order-$N$
relation and the residual real torsion factor is trivial, whereas for
even $N$ a nontrivial sign may remain. In the latter case,
\eqref{eq:abelianFGKphaseKR} reduces to
\begin{align}
 c_{\rm KR}^{[\mu,\omega],[\mu',\omega']}
=
(-1)^{\epsilon(\mu\wedge\omega'+\mu'\wedge\omega)}
\exp\{
i\pi\left(
B(\mu,\omega')-B(\omega,\mu')
\right)
\}.
\label{eq:abelianFGKphaseKRreal}
\end{align}
This constitutes the additional
topological contribution associated with the twisted sectors, which we identify below with the abelianization of the FGK cocycle of the WZW model.

\subsection{Abelianization of the WZ amplitude and the FGK cocycle}

We now return to the WZW model and show that, upon restriction to the maximal torus, the WZ amplitude obeys precisely the generalized abelian PW formula derived above. In this description, the WZ amplitude is represented by a constant $B$-field together with flat-gerbe data encoding the FGK cocycle.

Consider the maximal-torus-valued maps
\begin{align}
g_1
=
e^{
it(m+\mu)+is(w+\omega)
},\quad 
g_2
=
e^{
it(m'+\mu')+is(w'+\omega')
},
\end{align}
with $m,m',w,w'\in Q^\vee$ and
$\mu,\mu',\omega,\omega'\in\Lambda^\vee/Q^\vee\simeq\mathcal C$.
Applying the generalized PW formula~\eqref{eq:generalizedPW}, the PW cocycle becomes
\begin{align}
\Phi_{\rm WZ}(g_1,g_2)
&=
e^{
-\pi i\kappa
\left[
(m+\mu,w'+\omega')
-
(w+\omega,m'+\mu')
\right]
}.
\end{align}
Combining this with the level dependent part of the FGK cocycle gives
\begin{align}
\Phi_{\rm WZ}(g_1,g_2)
e^{
\pi i\kappa
\left(
(\mu,\omega')-(\omega,\mu')
\right)
}
= 
e^{
-\pi i\kappa
\left[
(m,w')-(w,m')
+(m,\omega')-(\omega,m')
+(\mu,w')-(w,\mu')
\right]
}.
\label{eq:PW-FGK-combined}
\end{align}
All the pairings remaining in~\eqref{eq:PW-FGK-combined} contain at least one coroot and are therefore integer-valued. Their contribution to the phase may therefore be represented by the $B$-field
$B=\kappa\mathsf B$, with $\mathsf B\equiv\mathsf C\pmod 2$. A representative
of $\mathsf B$ is
\begin{align}
\mathsf B_{ij}
=
\begin{cases}
\mathsf C_{ij}, & i<j,\\
0, & i=j,\\
-\mathsf C_{ij}, & i>j,
\end{cases}
\qquad
\mathsf C_{ij}=(\alpha_i^\vee,\alpha_j^\vee).
\end{align}
This is precisely the antisymmetric $B$-field datum of the distinguished
Narain point specified by $(Q^\vee,\kappa\mathsf C,\kappa\mathsf B)$ in the
abelianization of the simply connected WZW model~\cite{Lu_wzwlocalization2026}.
Equation~\eqref{eq:PW-FGK-combined} can thus be written as
\begin{align}
\Phi_{\rm WZ}(g_1,g_2)
e^{\pi i\kappa\left((\mu,\omega')-(\omega,\mu')\right)}
=
e^{-\pi i\kappa\left[
\mathsf B(m+\mu,w'+\omega')
-\mathsf B(w+\omega,m'+\mu')
\right]}
e^{\pi i\kappa\left[
\mathsf B(\mu,\omega')
-\mathsf B(\omega,\mu')
\right]}.
\end{align}
With $L=Q^\vee$, consider the $\mathcal C$-twisted maps into $T^d=\mathbb R^d/(2\pi L)$, with twists represented by $\mu,\omega\in\Lambda^\vee/Q^\vee$,
\begin{align}
X(t,s)
= 
t(m+\mu)+s(w+\omega), \quad 
X'(t,s)
&= 
t(m'+\mu')+s(w'+\omega'),
\end{align}
the first factor becomes
\begin{align}
\Phi_{\rm KR}(X,X')=\exp\{
-\frac{i}{4\pi}
\int_{T^2}
 (\kappa \mathsf B)_{ij}\,
dX^i\wedge dX'^j
\},
\end{align}
while the second is
\begin{align}
\exp\{
i\pi
\left[
(\kappa \mathsf B)(\mu,\omega')
-
(\kappa \mathsf B)(\omega,\mu')
\right]
\}.
\end{align}
Restoring the discrete torsion part of the FGK cocycle, the generalized WZW PW formula restricted to the maximal torus becomes
\begin{align}
A_{\rm WZ}[g_1g_2]
=
c_{\rm KR}^{[\mu,\omega],[\mu',\omega']} 
\Phi_{\rm KR}(X,X')
A_{\rm WZ}[g_1]A_{\rm WZ}[g_2].
\end{align}
This is precisely the generalized abelian PW formula~\eqref{eq:generalized-abelian-PW}, upon identifying $A_{\rm WZ}$ with $A_{\rm KR}$ and setting
\begin{align}
 L=Q^\vee,\quad
 B=\kappa\mathsf B,\quad
 a_B=(-1)^\epsilon.
\end{align}
Thus the WZ amplitude abelianizes to the $B$-field of the same distinguished Narain point obtained in the simply connected case, while the additional global information carried by the FGK cocycle is encoded in the corresponding flat-gerbe data.

\subsection{Siegel--Narain representation of the SWZW partition function}

The abelianization above allows us to identify the lattice sum appearing in each topological sector of the SWZW partition function with a generalized Siegel–Narain theta function. Recall that
\begin{align}
    Z_{{\rm SWZW}}
    =
    \sum_{\sigma \in W}
    \det(\sigma)\,
    \mathcal{Z}^\sigma, \quad     \mathcal{Z}
    =
    \frac{1}{\lvert \mathcal{C} \rvert}
    \sum_{\mu,\omega \in \mathcal{C}}
    \mathcal{E}_{\mu,\omega}\,
    \mathcal{Z}_{\mu,\omega}.
\end{align}
where $\mathcal Z_{\mu,\omega}$ and $\mathcal{E}_{\mu,\omega}$ are given by~\eqref{eq:EandZ}.
In particular, $\mathcal E_{\mu,\omega}$ captures the level-independent
part of $\Phi_{\mu,\omega}$ arising from the topological contributions
in the path integral,
\begin{align}
\mathcal E_{\mu,\omega}
=
(-1)^{\epsilon\mu\wedge\omega}
(-1)^{h^\vee
\left(
(\mu,\omega)+\mu^2+\omega^2
\right)}.
\end{align}
To exhibit the orbifold structure, we note that the modular
transformations of $\mathcal Z_{\mu,\omega}$ take the simple form
\begin{align}
\mathcal Z_{\mu,\omega}\big|_{T}
=
\mathcal Z_{\mu-\omega,\omega}, \quad 
\mathcal Z_{\mu,\omega}\big|_{S}
=
\mathcal Z_{\omega,-\mu},
\end{align}
while
\begin{align}
\mathcal E_{\mu-\omega,\omega}
=
\mathcal E_{\mu,\omega},\quad
\mathcal E_{\omega,-\mu} =
\mathcal E_{\mu,\omega}.
\end{align}
Hence the sum over $\mu, \omega \in \mathcal{C}$ is modular invariant and has the
standard structure of a simple-current orbifold: $\omega$ labels the
twisted sector, $\mu$ implements the orbifold projection, and
$\mathcal E_{\mu,\omega}$ supplies the discrete torsion phases.

This orbifold structure becomes explicit in the lattice realization of $\mathcal Z_{\mu,\omega}$.  As explained in~\cite{Lu_wzwlocalization2026}, the untwisted sector is described
by the Narain lattice
\begin{align}
    \Gamma= \{p=[p_L; p_R] \in \frac{1}{\kappa} P \times \frac{1}{\kappa} P: p_R - p_L \in  Q^\vee \}  . 
\end{align}
The twisted contribution can be written as
\begin{align}
\mathcal{Z}_{\mu,\omega}
= 2^r\, C_\kappa \, e^{\pi i\kappa(\mu,\omega)}
\sum_{\lambda\in P}
\sum_{\gamma\in \qqv}
 e^{2\pi i(\lambda,\mu)}
q^{\frac{\lambda^2}{2\kappa}}
\bar{q}^{\frac{1}{2\kappa}
\left(
\lambda+\kappa(\gamma+\omega)
\right)^2}
e^{2 \pi i (\lambda, u )}
e^{-2 \pi i ( \lambda+\kappa(\gamma+\omega), \bar v)},
\end{align}
with $ q=e^{2\pi i \tau}$, $\bar q = e^{-2\pi i \bar \tau}$. 
The summation variables naturally combine into momenta belonging to
the shifted Lorentzian lattice
\begin{align}
    \Gamma^\omega=\Gamma+\delta_\omega,
\quad
\delta_\omega=[0;\omega],
\end{align}
or equivalently,
\begin{align}
    \Gamma^\omega
=
\{
p=[p_L;p_R]\in
\frac{1}{\kappa}P\times
\frac{1}{\kappa}P:
p_R-p_L\in
Q^\vee+\omega
\},
\end{align}
with
\begin{align}
p_L = \frac{1}{\kappa}\lambda, \quad p_R = \frac{1}{\kappa }\lambda +  w + \omega  ,\quad \lambda \in P, \quad w \in Q^\vee.
\end{align}
Thus $\omega$ determines the shift of the Narain lattice.
Using the Lorentzian pairing
\begin{align}
\langle [x_L;x_R],[y_L;y_R]\rangle
=
\kappa (x_L,y_L)- \kappa (x_R,y_R),
\end{align}
and $\kappa(w,\mu)\in\mathbb Z$, we have
\begin{align}
e^{\pi i\langle\delta_\mu,\delta_\omega\rangle}
e^{-2\pi i\langle p,\delta_\mu\rangle}
=
e^{\pi i\kappa(\mu,\omega)}
e^{2\pi i(\lambda,\mu)}.
\end{align}
Thus $\delta_\omega$ determines the lattice shift, while $\delta_\mu $ implements the insertion associated with the orbifold projection.
The twisted contribution therefore takes the generalized Siegel–Narain form
\begin{align}
\label{eq:twistedpartfuncSN}
    \mathcal{\mathcal{Z}}_{\mu, \omega} 
     & =2^r\,  C_\kappa e^{\pi i \langle \delta_{\mu}, \delta_\omega \rangle } \sum_{p \in \Gamma + \delta_\omega }
     e^{-2\pi i \langle p, \delta_\mu \rangle} 
     e^{\pi i \tau \langle p, p \rangle_+  + \pi i \bar \tau \langle p, p \rangle_- }
     e^{2 \pi i \langle p, [u;\bar v] \rangle }\\
     & =   2^r\, \mathcal{S}^{\delta_{\mu}, \delta_{\omega}}_{\Gamma}(\tau, \bar \tau, [u;\bar v] ).
\end{align}
Since $\Gamma$ is even and unimodular, the modular transformations of the characteristics~\eqref{eq:SN-characteristics-T} and~\eqref{eq:SN-characteristics-S} reproduce precisely those of the topological sectors above
\begin{align}
T:\quad
(\delta_\mu,\delta_\omega)
&\longmapsto
(\delta_\mu-\delta_\omega,\delta_\omega),
\\
S:\quad
(\delta_\mu,\delta_\omega)
&\longmapsto
(\delta_\omega,-\delta_\mu).
\end{align}
The corresponding modular weight is $(r/2,r/2)$. Thus the generalized Siegel–Narain representation makes manifest the modular permutation of the twisted sectors. The full sum $\mathcal Z$ realizes the corresponding simple-current orbifold of the supersymmetric Narain theory, with $\mathcal E_{\mu,\omega}$ encoding the discrete torsion phases.

\section{Particle limit and generalized Frenkel's formula}
\label{sec:frenkel}

In the particle (large-$k$) limit, we set~\cite{Lu_wzwlocalization2026}
\begin{align}
    \tau = \frac{i\beta}{L}, \quad k = \frac{2\pi }{L}
\end{align}
and then take $L \to 0$. Then the spatial circle shrinks while the Euclidean time circle of complex length $\beta$ remains finite. The torus therefore reduces to a circle with a circumference of $\mathrm{Re}\, \beta$, and the theory becomes supersymmetric quantum mechanics on the compact Lie group $G$.  We will show that the resulting partition function becomes a generalized version of Frenkel's trace formula, valid also for non-simply connected groups.

\subsection{Particle limit in Hamiltonian formalism}

We now consider the Hamiltonian formulation of the SWZW partition function. Since the prefactor $C_\kappa$ has a finite limit given by
\begin{align*}
\lim_{L\to0}C_\kappa(u,v)
=
\exp\{
\frac{2\pi^2 i}{\beta}
\left(
(u,\operatorname{Im}u)
+
(\bar v,\operatorname{Im} \bar v)
\right)
\}.
\end{align*}
we will factor it out for convenience. We therefore consider
\begin{align}
  2^{-r}C_\kappa^{-1} Z_{\text{SWZW}} = \frac{1}{\lvert \mathcal{C} \rvert}
\sum_{\mu,\omega \in \mathcal{C}} \sum_{\lambda \in P^k_+}
(\pm 1)^{\mu \wedge \omega}
e^{2\pi i(\lambda,\mu)
+\pi ik(\mu,\omega)} N_{\lambda+\rho}
\bar{N}_{ s_\omega(\lambda) + \rho +\kappa\omega}.
\end{align}
Consider
\begin{align*}
N_{\lambda + \rho+ \kappa \omega,\kappa}(\tau, u)
 = &  \sum_{\sigma \in W} \det(\sigma) \Theta_{\sigma(\lambda + \rho + \kappa\omega), \kappa}(\tau, u) =   \sum_{\sigma \in W} \det(\sigma) \sum_{\gamma \in Q^\vee} f_{\sigma,\gamma+\omega}, 
\end{align*}
where
\begin{align*}
    f_{\sigma,\gamma} = \exp\{\frac{\pi i \tau}{\kappa} (\lambda + \rho + \kappa \gamma)^2+2\pi i\left(\lambda + \rho + \kappa \gamma,\sigma(u)\right)\}.
\end{align*}
In the $L \to 0$ limit, $f_{\sigma,\gamma + \omega}$ has a finite nonzero limit only when $\gamma + \omega=0$ which can occur only for $\gamma = 0 $ and $\omega=0$, whereas the other sectors are exponentially suppressed by $\exp[-2\pi^2 (\mathrm{Re}\,\beta/L^2)(\gamma+\omega)^2]$. 
Hence 
\begin{align}
    \lim_{L\to0} f_{\sigma,\gamma+ \omega} 
& = \delta_{\omega,0} \delta_{\gamma,0}
\exp\left(
-\frac{\beta}{2}(\lambda+\rho, \lambda+\rho)
+
2\pi i\bigl(\sigma(\lambda+\rho),u\bigr)
\right)
\end{align}
Hence 
\begin{align}
\lim_{L\to 0}
N_{\lambda+\rho+\kappa\omega,\kappa}
&=
\delta_{\omega,0}\,
e^{-\frac{\beta}{2}(\lambda+\rho)^2}
J_{\lambda+\rho},
\end{align}
We thus obtain the particle-limit formula
\begin{align}
\label{eq:particleLimitHamil}
\lim_{L\to0}2^{-r}C_\kappa^{-1} Z_{\rm SWZW}
&=
\frac{1}{|\mathcal C|}
\sum_{\mu\in\mathcal C}
\sum_{\lambda\in P_+}
e^{2\pi i(\lambda,\mu)}
e^{-\beta(\lambda+\rho)^2}
J_{\lambda+\rho}\bar J_{\lambda+\rho}
\nonumber\\
&=
\sum_{\lambda\in P_+}
{\bf 1}_{\lambda\in\Lambda}
e^{-\beta(\lambda+\rho)^2}
J_{\lambda+\rho}\bar J_{\lambda+\rho}
=
\sum_{\lambda\in\Lambda_+}
e^{-\beta(\lambda+\rho)^2}
J_{\lambda+\rho}\bar J_{\lambda+\rho},
\end{align}
where $\Lambda_+=\Lambda\cap P_+$ denotes the set of dominant character
weights. Here we have used the discrete Fourier projection
\begin{align}
\frac{1}{|\mathcal C|}
\sum_{\mu\in\mathcal C}
e^{2\pi i(\lambda,\mu)}
&=
{\bf 1}_{\lambda\in\Lambda},
\end{align}
which projects the weight lattice $P$ onto the character lattice $\Lambda$.

\subsection{Particle limit in localization formula}

On the other hand, the localization formula gives
\begin{align}
  2^{-r}C_\kappa^{-1} Z_{\text{SWZW}}
&= 
{\rm Vol}(\qqv)
\left(
\frac{\kappa}{2\tau_2}
\right)^{\frac{r}{2}}
\sum_{\sigma\in W}
\det(\sigma)
\frac{1}{\lvert\mathcal{C}\rvert}
\sum_{\mu,\omega\in\mathcal{C}}
\Phi_{\mu,\omega}
\sum_{m,w\in \qqv}
h^\sigma_{m+\mu,w+\omega}
\end{align}
where $\Phi_{\mu,\omega}$ is given in~\eqref{eq:Phi}, and
\begin{align}
h_{m,w}
& = 
e^{-\pi i \kappa (m, w)} \exp\{-\frac{\pi \kappa}{2 \tau_2}\left( (m - \tau w  + 2 u, m - \bar{\tau}w   - 2 \bar{v}))    +  (u + \bar{v})^2 \right)\}.
\end{align}

Similarly, $h_{m+\mu,w+\omega}$ has finite nonzero limit under $L\to0$ only when $w+\omega =0$ which again requires $w = 0$, $\omega =0 $, all other sectors are suppressed by $\exp[
-\pi^2(\mathrm{Re}\beta /L^2)(w+\omega)^2]$. Hence,
\begin{align}
    \lim_{L\to0} h_{m+\mu,w+\omega}
& =  \delta_{w,0} \delta_{\omega,0}
e^{-\frac{\pi^2}{\beta}\left( (m   + 2 u, m   - 2 \bar{v}))    +  (u + \bar{v})^2 \right)} 
= \delta_{w,0} \delta_{\omega,0}
e^{-\frac{\pi^2}{\beta}\left(m+u-\bar v \right)^2}.
\end{align}
Thus the localization formula reduces to
\begin{align}
\label{eq:particleLimitLocal}
     \lim_{L\to0}2^{-r}C_\kappa^{-1} Z_{\text{SWZW}} & = \left(\frac{\pi}{\beta}\right)^{r/2}
\text{Vol}\!\left(Q^\vee\right)
\sum_{\sigma\in W}
\det(\sigma) \frac{1}{\lvert\mathcal{C}\rvert}
\sum_{\mu\in\mathcal{C}}
\sum_{m \in Q^\vee + \mu } (-1)^{h^\vee \mu^2} e^{-\frac{\pi^2}{\beta}\left(m+\sigma(u) - \bar v \right)^2} \nonumber \\
& = \left(\frac{\pi}{\beta}\right)^{r/2}
{\rm Vol}\left(\Lambda^\vee\right)
\sum_{\sigma\in W}
\det(\sigma)
\sum_{m \in \Lambda^\vee}
e^{2\pi i(\rho,m)}
e^{-\frac{\pi^2}{\beta}\left(m+\sigma(u) - \bar v \right)^2}.
\end{align}
In particular, in the particle limit the phase $\Phi$ reduces to
\begin{align}
\Phi_{\mu,0}
=
(-1)^{h^\vee\mu^2}
=
(-1)^{(2\rho,\mu)},
\quad
\mu\in\Lambda^\vee/Q^\vee\simeq\mathcal C.
\end{align}
Equivalently, this phase can be written as
\begin{align}
\nu_m
 = (-1)^{(2\rho,m)} = e^{2\pi i(\rho,m)},
\quad
m\in\Lambda^\vee,
\end{align}
which is trivial on $Q^\vee$ and therefore descends to
$\Lambda^\vee/Q^\vee$.

The particle limit naturally yields the partition function of $\mathcal N=2$ supersymmetric quantum mechanics on the group manifold $G$. The two real supercharges descend from the $\mathcal N=(1,1)$
supersymmetry of the SWZW model, while the two Majorana-Weyl fermions reduce
to adjoint-valued one-dimensional Majorana fermions associated with the left-
and right-invariant frames of $G$. This differs from the formulation
of Choi and Takhtajan mainly in the treatment of the fermionic sector:
here the fermions descend directly from the SWZW model and retain the
$G_L\times G_R$ origin of the two chiral sectors. For simply connected $G$, the resulting partition function reduces to the particle-limit formula obtained in~\cite{Lu_wzwlocalization2026}.

\subsection{Generalized Frenkel's formula}
Equating~\eqref{eq:particleLimitHamil} and~\eqref{eq:particleLimitLocal} yields the generalized Frenkel trace formula
\begin{align} 
\label{eq:Frenkel}
\sum_{\lambda\in \Lambda_+}
\chi_\lambda(u)
\chi_\lambda(-\bar v) e^{-\beta c_2(\lambda)}
& = 
\frac{
\left(\frac{\pi}{\beta}\right)^{r/2}
{\rm Vol}(\Lambda^\vee)
e^{\beta(\rho,\rho)}
}{
J_\rho(u)
J_\rho( - \bar v)
} \nonumber \\
& \hspace{2cm } \times 
\sum_{\sigma\in W}
\det(\sigma)
\sum_{m\in \Lambda^\vee} e^{2\pi i(\rho,m)}
e^{
-\frac{\pi^2}{\beta}
\left(
m+ \sigma(u) - \bar v
\right)^2
},
\end{align}
which expresses the heat-kernel trace on the connected compact Lie group $G$ as a sum over the cocharacter lattice $\Lambda^\vee$.  For simply connected $G$, $\Lambda=P$ and $\Lambda^\vee=Q^\vee$, so the sign $ \nu_m = (-1)^{(2\rho,m)}$
is trivial and the formula reduces to the classical Frenkel trace formula; for non-simply connected groups, the sign $\nu_m$ depends only on the class of $m$ in $\Lambda^\vee/Q^\vee\cong\pi_1(G)$ and encodes the global topology of the target group.

We now give a direct verification of the generalized Frenkel trace formula. First, we start from the heat kernel trace side.  Using the Weyl character formula and expanding the Weyl numerators successively, the sum over dominant character weights is converted into a sum over the character lattice $\Lambda$, 
\begin{align}
    \sum_{\lambda\in\Lambda_+}
\chi_{\lambda}(u)
\chi_{\lambda}(-\bar v)e^{-\beta c_2(\lambda)} 
& = \frac{e^{\beta(\rho,\rho)}}{J_\rho(u)J_\rho(-\bar v)}\sum_{\lambda \in \Lambda} e^{2\pi i (\lambda+\rho,-\bar v)}
J_{\lambda+\rho}(u)e^{-\beta(\lambda+\rho)^2} \\
& =  \frac{e^{\beta(\rho,\rho)}}{J_\rho(u)J_\rho(-\bar v)} \sum_{\sigma\in W } \det(\sigma) \sum_{\lambda \in \Lambda} e^{2\pi i (\lambda+\rho,\sigma(u) -\bar v)} e^{-\beta(\lambda+\rho)^2}
\end{align}
Applying Poisson resummation on the character lattice $\Lambda$, whose dual lattice is the cocharacter lattice $\Lambda^\vee$, gives
\begin{align}
\left(\frac{\pi}{\beta}\right)^{r/2}
{\rm Vol}(\Lambda^\vee)\frac{e^{\beta(\rho,\rho)}}{J_\rho(u)J_\rho(-\bar v)}\sum_{\sigma\in W}
\det(\sigma)
\sum_{m\in \Lambda^\vee} e^{-2\pi i(\rho,m)}
e^{
-\frac{\pi^2}{\beta}
\left(
m+ \sigma(u) -\bar  v
\right)^2
} .
\end{align}
Since $e^{-2\pi i(\rho,m)} = e^{2\pi i(\rho,m)}$, this reproduces the right-hand side of the generalized Frenkel trace formula.

\subsection{One-dimensional fermion global anomaly}

In the localization approach to supersymmetric quantum mechanics on a
compact Lie group $G$, the sign
\begin{align}
\nu_m=(-1)^{(2\rho,m)},
\quad
m\in\Lambda^\vee,
\end{align}
appearing in the generalized Frenkel trace formula admits a natural
interpretation as a global anomaly of the adjoint fermion.

One way to see this is through the regularized fermion Pfaffian, represented
by the Weyl denominator~\cite{choiT25}
\begin{align}
    J_\rho(u)= \prod_{\alpha\in R_+} (e^{\pi i (u,\alpha)} - e^{-\pi i (u,\alpha)}).
\end{align} 
Under a large gauge transformation
of the background gauge field,
\begin{align}
J_\rho(u+m)
=
(-1)^{\sum_{\alpha>0}(\alpha,m)} J_\rho(u)
=
\nu_m J_\rho(u),
\quad
m\in\Lambda^\vee.
\end{align}
Since $\nu_m=1$ for $m\in Q^\vee$, the phase depends only on
$[m]\in\Lambda^\vee/Q^\vee\simeq\pi_1(G)$. Thus $\nu_m$ measures the
relative phase of the Pfaffian under a large gauge transformation and may
be interpreted as the corresponding global fermion anomaly.

More intrinsically, this anomaly may be described as the holonomy of
the fermion Pfaffian line under a large gauge transformation~\cite{Witten2015FermionPI,WittenYonekura2019eta, Koizumi2021DimOne}, in close
analogy with the two-dimensional Pfaffian holonomy discussed above.
Decomposing the non-Cartan part of the adjoint representation into real
root planes, we write
\begin{align}
\nu_m
=
\prod_{\alpha\in R_+}\nu_m^\alpha,
\end{align}
where $\nu_m^\alpha$ is the relative Pfaffian holonomy associated with
the real two-dimensional plane
$\mathfrak g_\alpha\oplus\mathfrak g_{-\alpha}$.

For each root plane, the large gauge transformation
$[m]\in\Lambda^\vee/Q^\vee$ defines a loop in the space of background
gauge fields. The holonomy of the corresponding real Pfaffian line is
given by the mod-$2$ spectral flow, equivalently by the mod-$2$ index of
the Dirac operator on the associated two-dimensional mapping torus,
\begin{align}
\nu_m^\alpha
=
(-1)^{
{\rm Ind}_2(\mathcal D^\alpha_m)
}.
\end{align}
The large gauge transformation changes the effective spin structure
seen by the root plane according to
\begin{align}
{\rm P}_m
=
{\rm P}+(\alpha,m)
\pmod 2,
\end{align}
where ${\rm P}$ denotes the periodic spin structure. Consequently,
\begin{align}
{\rm Ind}_2(\mathcal D^\alpha_m)
=
(\alpha,m)
\pmod 2,
\end{align}
and hence
\begin{align}
\nu_m^\alpha
=
(-1)^{(\alpha,m)}.
\end{align}
Multiplying over the positive roots gives
\begin{align}
\nu_m
=
(-1)^{\sum_{\alpha>0}(\alpha,m)}
=
(-1)^{(2\rho,m)}.
\end{align}
Thus the Weyl-denominator regularization and the mod-$2$ index compute
the same holonomy of the fermion Pfaffian line bundles.

\section{Summary and future directions}
\label{sec:summary}

In this work, we have studied the SWZW model with a non-simply connected target group, with particular emphasis on the global topological data that enter its localization formula and their relation to an abelian Narain description. Compared with the simply connected case, the essential new ingredients arise from the nontrivial fundamental group of the target and appear both in the classical WZ amplitude and in the global anomaly of the Majorana--Weyl fermion Pfaffian.

Specifically, we have clarified the role of the FGK cocycle in the WZ amplitude and its interpretation in terms of gerbe holonomy. For field configurations restricted to the maximal torus, the Wess--Zumino term admits an abelian description in which its local contribution is represented by a constant $B$-field, while the remaining torsion information is encoded by a flat gerbe. The latter is responsible for the discrete FGK phases associated with the nontrivial winding sectors. This gives a direct geometric interpretation of the additional phases that appear when passing from the simply connected theory to a quotient target.

Furthermore, motivated by the global-anomaly viewpoint in~\cite{Witten1985Global, FreedVafa87Global}, we have identified the global anomaly phases
of the fermionic Pfaffian with relative Rochlin invariants~\cite{LeeMillerWeintraub1988}. For the torus, these invariants can be expressed in terms of Arf invariants and the quadratic refinements
associated with the choice of spin structure. 
In this way, the winding-dependent signs arising from the fermionic
one-loop determinant acquire a purely topological interpretation. Together with the FGK cocycle of the WZ amplitude, they provide the additional discrete phases entering the localization formula for non-simply connected targets. The resulting combination makes explicit the interplay between fermionic and WZ global data under abelianization.

A further outcome of our analysis is a generalized Narain realization of the SWZW partition function. After abelianization, the sum over topological sectors can be reorganized as a simple-current orbifold of the level-$\kappa$ theta-function theory. The different orbifold sectors correspond to shifted Narain lattices, and their partition functions are naturally expressed as generalized Siegel--Narain theta functions. The phases inherited from the FGK cocycle and the fermion global anomaly combine into the discrete phases accompanying the corresponding orbifold sectors. Thus, after localization, the global geometric data of the original nonabelian WZW model are translated into the shift and discrete torsion data of an abelian lattice theory.

Several natural extensions of these results remain to be explored. One direction
is to formulate a Duistermaat--Heckman (DH) formula directly for non-simply
connected targets. In the simply connected setting, supersymmetric localization
provides a close relation between the WZW path integral and the corresponding
equivariant localization formula~\cite{Wendt2001WZW}. It would be interesting
to understand how this picture is modified by the additional topological sectors
and, in particular, how the FGK and fermionic phases should be incorporated into
a generalized DH formula. Such a formulation may provide a more intrinsic
localization interpretation of the discrete topological data found in the
present work.

A closely related direction is to explore the duality web relating quantum
mechanics on a group manifold, WZW theory, two-dimensional Yang--Mills
theory~\cite{witten912dYM,witten922dYMrevisited}, and three-dimensional
Chern--Simons theory~\cite{Witten1988CSJones,elitzurMooreSeibergSchwimmer89}.
It would be interesting to understand how the global topology of a non-simply
connected target modifies these correspondences and their associated localization
structures. In particular, when the level satisfies the CS admissibility
condition, the simple-current extension of the WZW chiral algebra should admit
a direct three-dimensional interpretation in terms of Chern--Simons theory with
the corresponding non-simply connected gauge group. The Chern--Simons path
integral on $T^2\times I$, viewed as a propagator on the Hilbert space associated
with $T^2$, provides a natural setting for relating the diagonal pairing of
extended conformal blocks to the non-simply connected WZW partition function.
This perspective may also be compared with nonabelian DH equivariant localization
of Chern--Simons theory~\cite{BeasleyWitten2005} and its
abelianization~\cite{BlauThompson2006S1bundle}. More generally, one may ask how
these localization approaches are related across the duality web and whether the
discrete sectors and phases identified here admit a unified interpretation within
the corresponding localization formulas.

From the perspective of the abelianized theory, another direction is to
understand how these structures behave under deformations and dualities. The
shifted Narain description obtained here identifies the topological sectors with
shifted lattice sectors carrying the associated discrete phases. This suggests
studying their behavior under exactly marginal current--current deformations of
WZW models~\cite{Forste:2003km} and under $T$-duality
transformations~\cite{Kiritsis:1993ju,Gaberdiel:1995mx}. In particular, it
would be interesting to determine how the shifts and the phases arising from
the FGK cocycle and the fermion global anomaly transform under these
operations. More broadly, this may provide a localization perspective on
moduli-space averaging in deformed WZW and Narain
theories~\cite{Maloney:2020nni,Afkhami-Jeddi:2020ezh,Dong:2021wot}, as well
as on distinguished lattice theories such as code
CFTs~\cite{Dymarsky:2020qom}.

Another direction is to extend the analysis beyond the torus to more general
worldsheet topologies. On higher-genus closed worldsheets, the classification of
topological sectors and the dependence on spin structures become substantially
richer. One expects corresponding higher-genus generalizations of the abelianization formula and of the shifted Narain description, together with a more intricate realization of the FGK cocycle and fermion global anomaly. Understanding these extensions may help clarify more generally how the global topology of the nonabelian target is encoded in the abelian theory obtained through localization.

A complementary direction is to consider worldsheets with boundary. In this setting, the WZ gerbe requires appropriate boundary data, naturally bringing the localization problem into contact with boundary conformal field theory and D-branes~\cite{GawedzkiReis2002braneGerbe}. Recent work has also related the higher Berry connection on boundary conformal manifolds to the NS–NS $B$-field in the D-brane interpretation~\cite{ChoiHKKusukiO2025BerryBCFT}. For WZW models, D-brane charges are closely related to twisted $K$-theory, while for non-simply connected targets the boundary-state structure reflects the simple-current orbits and fixed-point phenomena that also underlie the bulk theory~\cite{GaberdielGannon2004Dbrane}. It would be interesting to determine whether supersymmetric localization can be extended to such worldsheets and how the topological sectors, FGK cocycles, and fermion global anomalies appearing in the closed-string analysis are encoded in boundary states and open-string amplitudes~\cite{FreedWitten1999AnomalyDbrane, Kapustin:1999DbraneBfield}. Such an extension may provide a path-integral perspective on the relation between SWZW models, boundary conformal field theory, and twisted $K$-theory~\cite{FredenhagenSchomerus2000Ktheory, Gawedzki2004AbelianNonAbelian}.

\appendix

\section{Lie groups, global forms, and lattices}
\label{apx:lieGroup}

We begin by fixing our conventions for compact simple Lie groups and their
associated lattices. Let $\widetilde G$ be the simply connected compact
simple Lie group with Lie algebra $\mathfrak g$. We use the invariant bilinear form $(\cdot ,\cdot)$, normalized by $(\theta,\theta)=2$, to identify the Cartan sub-algebra $\mathfrak h$ with its dual $\mathfrak h^\ast$. A
general compact connected simple Lie group with Lie algebra $\mathfrak g$
is of the form
\begin{align}
G=\widetilde G/\mathcal C,
\quad
\mathcal C\subset Z(\widetilde G),
\end{align}
where $\mathcal C$ is a subgroup of the center. Its fundamental group is
therefore $\pi_1(G)\simeq \mathcal C$.
The adjoint group corresponds to the quotient by the full center,
\begin{align}
G_{\rm ad}=\widetilde G/Z(\widetilde G).
\end{align}

\begin{figure}[H]
    \centering

    \begin{tikzpicture}[
        solid line/.style={line width=0.8pt},
        dashed line/.style={dashed, line width=0.6pt},
        vertex/.style={fill=black, circle, inner sep=2.5pt}
    ]

        \def\k{0.6}
        \def\dx{2*\k}
        \def\dy{2*\k}
        \def\shift{3}

        \begin{scope}[xshift=-\shift cm]

            \node[vertex] (P)     at (0,\dy) {};
            \node[vertex] (Qvee)  at (0,-\dy) {};
            \node[vertex] (Q)     at (-\dx,0) {};
            \node[vertex] (Pvee)  at (\dx,0) {};

            \node[above=5pt] at (P)    {$P$};
            \node[below=5pt] at (Qvee) {$\qqv$};
            \node[left=5pt]  at (Q)    {$Q$};
            \node[right=5pt] at (Pvee) {$P^\vee$};

            \draw[solid line] (P) -- (Q);
            \draw[solid line] (P) -- (Pvee);
            \draw[solid line] (Q) -- (Qvee);
            \draw[solid line] (Pvee) -- (Qvee);

            \draw[dashed line] (Q) -- (Pvee);
            \draw[dashed line] (P) -- (Qvee);

        \end{scope}

        \begin{scope}[xshift=\shift cm, yshift= 0.2 cm]

            \def\dyTop{2.5*\k}
            \def\dyMid{1*\k}

            \node[vertex] (P2)          at (0,\dyTop) {};
            \node[vertex] (Qvee2)       at (0,-\dyTop) {};

            \node[vertex] (GammaG2)     at (-\dx,\dyMid) {};
            \node[vertex] (Pvee2)       at (\dx,\dyMid) {};

            \node[vertex] (Q2)          at (-\dx,-\dyMid) {};
            \node[vertex] (GammaGvee2)  at (\dx,-\dyMid) {};

            \node[above=5pt] at (P2)          {$P$};
            \node[below=5pt] at (Qvee2)       {$\qqv$};

            \node[left=5pt]  at (GammaG2)     {$\Lambda_G$};
            \node[right=5pt] at (Pvee2)       {$P^\vee$};

            \node[left=5pt]  at (Q2)          {$Q$};
            \node[right=5pt] at (GammaGvee2)  {$\Lambda_G^\vee$};

            \draw[solid line] (P2) -- (GammaG2);
            \draw[solid line] (P2) -- (Pvee2);

            \draw[solid line] (GammaG2) -- (Q2);
            \draw[solid line] (Pvee2) -- (GammaGvee2);

            \draw[solid line] (Q2) -- (Qvee2);
            \draw[solid line] (GammaGvee2) -- (Qvee2);

            \draw[dashed line] (GammaG2) -- (GammaGvee2);
            \draw[dashed line] (P2) -- (Qvee2);
            \draw[dashed line] (Q2) -- (Pvee2);

        \end{scope}
    \node at (-\shift,-2.5) {\textbf{(a)} };

    \node at (\shift,-2.5) {\textbf{(b)}}; 
    \end{tikzpicture}
   \caption{(a) Relations among the root, coroot, weight, and coweight lattices intrinsic to the Lie algebra $\mathfrak g$. (b) The same diagram augmented by the character and cocharacter lattices $\Lambda_G$ and $\Lambda_G^\vee$ associated with a general global form $G$. Solid lines indicate lattice inclusions, while dashed lines connect mutually dual lattices with respect to the invariant bilinear form.}
    \label{fig:lattice-hasse-diagrams}
\end{figure}
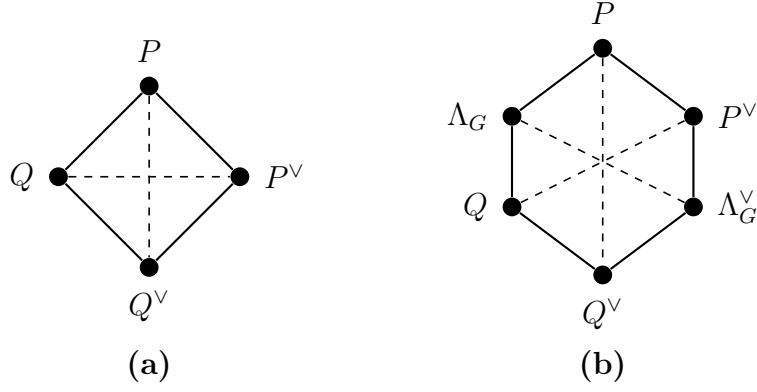

Let $Q$, $Q^\vee$, $P$, and $P^\vee$ denote the root, coroot, weight,
and coweight lattices, respectively. These lattices are intrinsic to
the Lie algebra $\mathfrak g$. For a choice of global form $G$, we
further introduce the character and cocharacter lattices
$\Lambda_G$ and $\Lambda_G^\vee$. When the global form is understood,
we abbreviate these as $\Lambda$ and $\Lambda^\vee$, respectively. They satisfy
\begin{align}
Q\subseteq \Lambda_G\subseteq P,
\quad
Q^\vee\subseteq \Lambda_G^\vee\subseteq P^\vee.
\end{align}
The simply connected and adjoint groups correspond to
\begin{align}
\Lambda_{\widetilde G}=P,
\quad
\Lambda_{G_{\rm ad}}=Q,
\end{align}
and similarly for the cocharacter lattices,
\begin{align}
\Lambda_{\widetilde G}^\vee=Q^\vee,
\quad
\Lambda_{G_{\rm ad}}^\vee=P^\vee.
\end{align}
The center and fundamental group can be expressed uniformly in terms
of the character and cocharacter lattices:
\begin{align}
Z(G)
&\simeq
\Lambda_G/Q
\simeq
P^\vee/\Lambda_G^\vee,
\\
\pi_1(G)
&\simeq
P/\Lambda_G
\simeq
\Lambda_G^\vee/Q^\vee.
\end{align}
The nontrivial classes in $P^\vee/Q^\vee$ may be represented by the
corresponding minuscule coweights associated with the special nodes of
the extended Dynkin diagram.  Accordingly, when discussing a subgroup
$\mathcal C\subset Z(\widetilde G)$, we choose representatives
$\mu,\omega\in P^\vee$ for its elements, with generators chosen among
these minuscule coweights.  We will often use the same symbols
$\mu,\omega$ for the corresponding classes modulo $Q^\vee$ when no
confusion can arise.
The relations among the various lattices are summarized in
Fig.~\ref{fig:lattice-hasse-diagrams}.

\section{WZW and Chern--Simons constraints}
\label{apx:constraints}

The WZW and Chern--Simons theories impose different quantization
conditions on the level $k$ for a non-simply connected group
$G=\widetilde G/\mathcal C$. We collect here the relevant
group-theoretic data and the resulting constraints used in the
subsequent appendices, and examine the admissibility of the shifted
level $\kappa=k+h^\vee$.

\subsection{WZW constraint}

The WZ amplitude is well defined when the level-$k$
WZ three-form defines an integral class in
$H^3(G,\mathbb Z)$. Equivalently, WZW admissibility of $k$ is the
condition for the existence of a bundle gerbe over $G$, whose curvature
is the level-$k$ WZ three-form~\cite{GawedzkiReis2002braneGerbe}. For the simply connected group this gives
$k\in\mathbb Z$. For a quotient $G=\widetilde G/\mathcal C$,
additional conditions arise from the integrality of the periods of the
WZ three-form.

For each cyclic quotient $\mathcal C$, we choose a generator represented
by a minuscule coweight $\gamma\in P^\vee$. The WZW condition is
\begin{align}
\frac{kN}{2}\gamma^2\in\mathbb Z.
\label{eq:WZW-cyclic-constraint}
\end{align}

For $D_{2r}$, the center is $\mathbb Z_2\times\mathbb Z_2$. We use the
two minuscule spinor coweights
\begin{align}
\gamma_s=\lambda_{2r-1}^\vee,
\quad
\gamma_c=\lambda_{2r}^\vee,
\end{align}
together with the vector class represented by $\gamma_v=\gamma_s+\gamma_c$. 
Thus
\begin{align}
\mathbb Z_2^s=\langle\gamma_s\rangle,
\quad
\mathbb Z_2^c=\langle\gamma_c\rangle,
\quad
\mathbb Z_2^v=\langle\gamma_v\rangle.
\end{align}
For the full quotient $\mathcal C=\mathbb Z_2\times\mathbb Z_2$,
the WZW condition may be imposed on the three nontrivial classes,
\begin{align}
k\gamma_s^2\in\mathbb Z,
\quad
k\gamma_c^2\in\mathbb Z,
\quad
k(\gamma_s+\gamma_c)^2\in\mathbb Z.
\label{eq:WZW-D2r-full}
\end{align}

The minuscule-coweight representatives and their relevant inner
products for the possible quotients are summarized in
Table~\ref{table:generator-data}.
\begin{table}[h]
    \centering
\begin{align*}
\renewcommand{\arraystretch}{1.4}
\begin{array}{c|c|c}
\hline
\mathfrak g & \mathcal C & \text{minuscule-coweight data}\\
\hline
A_{r-1}
& \mathbb Z_N,\ N\mid r
& \gamma^2=\frac{r(r-1)}{N^2}
\\
\hline
B_r
& \mathbb Z_2
& \gamma^2=1
\\
\hline
C_r
& \mathbb Z_2
& \gamma^2=\frac r2
\\
\hline
D_{2r+1}
&
\begin{array}{c}
\mathbb Z_2\\[1mm]
\mathbb Z_4
\end{array}
&
\begin{array}{c}
\gamma^2=1\\[1mm]
\gamma^2=\frac{2r+1}{4}
\end{array}
\\
\hline
D_{2r}
&
\begin{array}{c}
\mathbb Z_2^v\\[1mm]
\mathbb Z_2^{s,c}\\[1mm]
\mathbb Z_2\times\mathbb Z_2
\end{array}
&
\begin{array}{c}
\gamma_v^2=1\pmod{2\mathbb Z}\\[1mm]
\gamma_s^2=\gamma_c^2=\frac r2\\[1mm]
\gamma_s^2=\gamma_c^2=\frac r2,\quad
(\gamma_s,\gamma_c)=\frac{r-1}{2}
\end{array}
\\
\hline
E_6
& \mathbb Z_3
& \gamma^2=\frac43
\\
\hline
E_7
& \mathbb Z_2
& \gamma^2=\frac32
\\
\hline
\end{array}
\end{align*}
\caption{Generator data for the possible nontrivial quotients
$\mathcal C\subset Z(\widetilde G)$.}
\label{table:generator-data}
\end{table}

\subsection{Chern--Simons constraint.}

For simply connected target $G$, the 2d WZW theory is closely related
to 3d Chern--Simons (CS) theory at the same level $k$, involving the
same bundle gerbe over $G$. For a non-simply connected group, however,
the CS theory imposes stronger constraints on the level, corresponding
to the requirement that the level-$k$ WZW gerbe admit a multiplicative
structure needed to define the CS amplitude unambiguously~\cite{GawedzkiWaldorf2009PWMulGerb}. The obstruction to such a multiplicative structure is equivalently encoded by the FGK cocycle~\eqref{eq:FGKcocycle} appearing in the generalized PW formula. 

The CS constraints have a natural interpretation in terms of the simple
currents associated with $\mathcal C$. For $\omega\in\mathcal C$, the
corresponding simple current $J_\omega$ has conformal weight~\cite{MooreSeiberg1989Taming}
\begin{align}
h_{J_\omega}
=
\frac{
(k\omega,k\omega+2\rho)
}{
2(k+h^\vee)
}
=
\frac{k}{2}\omega^2
\pmod{\mathbb Z},
\end{align}
while the monodromy charge is
\begin{align}
Q_{J_\mu}(J_\omega)
=
h_{J_\mu}+h_{J_\omega}-h_{J_{\mu+\omega}}
=
-k(\mu,\omega)
\pmod{\mathbb Z}.
\end{align}
The associated Kreuzer--Schellekens bihomomorphism (KSB)
$\phi$ on $\mathcal C\times\mathcal C$ is~\cite{KreuzerSchellekens1993}
\begin{align}
\label{eq:KSB}
\phi_{\mu,\omega}
=
(-1)^{\epsilon \mu\wedge\omega}
e^{\pi i k(\mu,\omega)},
\end{align}
and satisfies
\begin{align}
\phi_{\mu,\mu}
=
e^{2\pi i h_{J_\mu}},
\quad
\phi_{\mu,\omega}\phi_{\omega,\mu}
=
e^{-2\pi i Q_{J_\mu}(J_\omega)}.
\end{align}
Here the factor $(-1)^{\epsilon\mu\wedge\omega}$ encodes the discrete torsion choice. The FGK cocycle is determined by KSB through
\begin{align}
    c_{\rm WZ}^{[\mu, \omega], [\mu', \omega']} = \frac{\phi_{\mu, \omega'}}{\phi_{\mu', \omega}}.
\end{align}
Consequently, triviality of the KSB is equivalent to triviality of the
FGK cocycle and hence to the absence of the multiplicative-gerbe
obstruction. CS admissibility can therefore be expressed as
\begin{align}
\phi_{\mu,\omega}=1,
\qquad
\mu,\omega\in\mathcal C.
\end{align}
This condition implies that the corresponding simple currents have
integer conformal spin and trivial mutual monodromy,
\begin{align}
h_{J_\mu}\in\mathbb Z,
\qquad
Q_{J_\mu}(J_\omega)\in\mathbb Z,
\end{align}
and hence generate a consistent simple-current extension of the WZW
chiral algebra~\cite{SchellekensYankielowicz1989Extended, Intriligator1989Bonus}.

For cyclic quotients $\mathcal C=\mathbb Z_N$, there is no
nontrivial discrete torsion choice, and the CS constraint reduces to
\begin{align}
\frac{k}{2}\omega^2\in\mathbb Z,
\quad
k(\mu,\omega)\in\mathbb Z,
\quad
\mu,\omega\in\mathcal C.
\label{eq:CS-general-constraint}
\end{align}

For $D_{2r}$ with
$\mathcal C=\mathbb Z_2\times\mathbb Z_2$, there is in addition a
discrete torsion choice $\epsilon\in\{0,1\}$. In terms of the generators $\gamma_s,\gamma_c$, CS
admissibility requires
\begin{align}
\frac{k}{2}\gamma_s^2\in\mathbb Z,
\quad
\frac{k}{2}\gamma_c^2\in\mathbb Z,
\quad
k(\gamma_s,\gamma_c)+\epsilon\in2\mathbb Z.
\label{eq:CS-D2r-full}
\end{align}
Combining the results of the preceding two subsections gives the
complete WZW and Chern--Simons constraints summarized in
Table~\ref{table:FGK}.
\begin{table}[h]
\centering
\begin{align*}
\renewcommand{\arraystretch}{1.6}
\begin{array}{ccccc}
\hline
\mathfrak g & \text{Center} & \mathcal C
& \text{WZW constraint on } k
& \text{CS constraint on } k
\\
\hline
A_{r-1}
& \mathbb Z_r
& \mathbb Z_N,\ N\mid r
& 2N\mid kr(r-1)
& 2N^2\mid kr(r-1)
\\
\hline
B_r
& \mathbb Z_2
& \mathbb Z_2
& -
& 2\mid k
\\
\hline
C_r
& \mathbb Z_2
& \mathbb Z_2
& 2\mid kr
& 4\mid kr
\\
\hline
D_{2r+1}
& \mathbb Z_4
&
\begin{array}{c}
\mathbb Z_2\\[1mm]
\mathbb Z_4
\end{array}
&
\begin{array}{c}
-\\[1mm]
2\mid k
\end{array}
&
\begin{array}{c}
2\mid k\\[1mm]
8\mid k
\end{array}
\\
\hline
D_{2r}
& \mathbb Z_2\times\mathbb Z_2
&
\begin{array}{c}
\mathbb Z_2^v\\[1mm]
\mathbb Z_2^s\\[1mm]
\mathbb Z_2^c\\[1mm]
\mathbb Z_2\times\mathbb Z_2
\end{array}
&
\begin{array}{c}
-\\[1mm]
2\mid kr\\[1mm]
2\mid kr\\[1mm]
2\mid k
\end{array}
&
\begin{array}{c}
2\mid k\\[1mm]
4\mid kr\\[1mm]
4\mid kr\\[1mm]
4\mid kr,\quad 4\mid(k+2\epsilon)
\end{array}
\\
\hline
E_6
& \mathbb Z_3
& \mathbb Z_3
& -
& 3\mid k
\\
\hline
E_7
& \mathbb Z_2
& \mathbb Z_2
& 2\mid k
& 4\mid k
\\
\hline
\end{array}
\end{align*}
\caption{WZW and CS constraints on the level $k$. A dash means that
integrality of $k$ imposes no additional restriction.}
\label{table:FGK}
\end{table}

\section{Explicit form of the phase $\Phi$}
\label{apx:phasePhi}

We collect here the explicit form of the phase
\begin{align}
\Phi_{\mu,\omega}
=
(-1)^{h^\vee(\mu^2+\omega^2)}
(-1)^{\epsilon \mu \wedge \omega}
\exp\{
\pi i k(\mu,\omega)
\}
\end{align}
for the possible global forms $G=\widetilde G/\mathcal C$.
For $\mathfrak g=G_2,F_4,E_8$, the center is trivial and hence
$\Phi=1$. All remaining quotients are cyclic except for the full
center of $D_{2r}$. Let $\mathcal C=\mathbb Z_N$ be generated by the coweight class
$\gamma$ specified in Appendix~\ref{apx:constraints}. Writing
\begin{align}
\mu=m\gamma,
\quad
\omega=n\gamma,
\quad
m,n\in\mathbb Z_N,
\end{align}
the discrete torsion contribution is trivial and
\begin{align}
\Phi^{\mathfrak g,\mathcal C}_{m,n}
=
(-1)^{h^\vee\gamma^2(m^2+n^2)}
\exp\{
\pi i k\gamma^2mn
\}.
\end{align}
Using the generator data in Table~\ref{table:generator-data}, this gives
\begin{align}
\Phi^{A_{r-1},\,\mathbb Z_N}_{m,n}
&=
(-1)^{
\frac{r^2(r-1)}{N^2}(m^2+n^2)
}
\exp\{
\frac{\pi i k r(r-1)}{N^2}mn
\},
\\
\Phi^{B_r,\,\mathbb Z_2}_{m,n}
&=
(-1)^{m+n+k mn} ,
\\
\Phi^{C_r,\,\mathbb Z_2}_{m,n}
&=
(-1)^{
\frac{r(r+1)}{2}(m^2+n^2)
}
\exp\{
\frac{\pi i kr}{2}mn
\} ,
\\
\Phi^{D_{2r+1},\,\mathbb Z_2}_{m,n}
&=
(-1)^{k mn} ,
\\
\Phi^{D_{2r+1},\,\mathbb Z_4}_{m,n}
&=
(-1)^{r(2r+1)(m^2+n^2)}
\exp\{
\frac{\pi i k(2r+1)}{4}mn
\} ,
\\
\Phi^{D_{2r},\,\mathbb Z_2^v}_{m,n}
&=
(-1)^{k mn},
\\
\Phi^{D_{2r},\,\mathbb Z_2^{s,c}}_{m,n}
&=
(-1)^{r(m+n)}
\exp\{
\frac{\pi i kr}{2}mn
\} , \\
\Phi^{E_6,\,\mathbb Z_3}_{m,n}
&=
\exp\{
\frac{4\pi i k}{3}mn
\},
\\
\Phi^{E_7,\,\mathbb Z_2}_{m,n}
&=
(-1)^{m+n}
\exp\{
\frac{3\pi i k}{2}mn
\},
\end{align}
The only noncyclic case is the full center
$\mathcal C=\mathbb Z_2\times\mathbb Z_2$ of $D_{2r}$.
Using the generators $\gamma_s,\gamma_c$ defined in
Appendix~\ref{apx:constraints}, we write
\begin{align}
\mu=m_s\gamma_s+m_c\gamma_c,
\quad
\omega=n_s\gamma_s+n_c\gamma_c,
\quad
m_s,m_c,n_s,n_c\in\mathbb Z_2.
\end{align}
The relevant bilinear and quadratic forms are
\begin{align}
(\mu,\omega)
&=
\frac12\left(
r(m_sn_s+m_cn_c)
+
(r-1)(m_sn_c+m_cn_s)
\right),
\\
\mu\wedge\omega
&=
m_sn_c-m_cn_s,
\\
h^\vee(\mu^2+\omega^2)
&\equiv
r(m_s+m_c+n_s+n_c)
\pmod2.
\end{align}
Including the discrete torsion contribution $(-1)^{\epsilon\,\mu\wedge\omega}$ with $\epsilon\in\{0,1\}$, we obtain
\begin{align}
\Phi^{D_{2r},\,\mathbb Z_2\times\mathbb Z_2}
_{(m_s,m_c),(n_s,n_c)}
&=
(-1)^{r(m_s+m_c+n_s+n_c)}
(-1)^{\epsilon(m_sn_c-m_cn_s)}
\nonumber\\
&\quad\times
\exp\{
\frac{\pi i k}{2}
\left(
r(m_sn_s+m_cn_c)
+
(r-1)(m_sn_c+m_cn_s)
\right)
\}.
\end{align}
Restriction to $\mathbb Z_2^v$, $\mathbb Z_2^s$, or
$\mathbb Z_2^c$ reproduces the corresponding cyclic expressions
above.

\section{Triviality of the fermion phase $\nu$}
\label{apx:phaseNu}

We determine when the fermionic phase
\begin{align}
\nu(\mu,\omega)
=
(-1)^{h^\vee\left((\mu,\omega)+\mu^2+\omega^2\right)}
\end{align}
is trivial for all $\mu,\omega\in\mathcal C$.

For $\mathcal C=\mathbb Z_N$, writing $\mu=m\gamma$, $\omega=n\gamma$, $
m,n\in\mathbb Z_N$, 
we obtain
\begin{align}
\nu(m,n)
=
(-1)^{h^\vee\gamma^2(mn+m^2+n^2)}.
\end{align}
Since $mn+m^2+n^2$ can be odd, $\nu$ is trivial precisely when
\begin{align}
h^\vee\gamma^2\in2\mathbb Z.
\end{align}
Using the generator data in Table~\ref{table:generator-data}, this
gives all cyclic cases summarized in
Table~\ref{table:nu-triviality}.

For
$\mathcal C=\mathbb Z_2\times\mathbb Z_2$ of $D_{2r}$, write
\begin{align}
\mu=m_s\gamma_s+m_c\gamma_c,
\quad
\omega=n_s\gamma_s+n_c\gamma_c,
\quad
m_s,m_c,n_s,n_c\in\mathbb Z_2.
\end{align}
Using the definition of $\nu$, its exponent becomes
\begin{align}
h^\vee\left((\mu,\omega)+\mu^2+\omega^2\right)
&\equiv
h^\vee\gamma_s^2
\left(m_sn_s+m_s+n_s\right)
+
h^\vee\gamma_c^2
\left(m_cn_c+m_c+n_c\right)
\nonumber\\
&\hspace{4.3cm}
+
h^\vee(\gamma_s,\gamma_c)
\left(m_sn_c+m_cn_s\right)
\pmod2.
\end{align}
Therefore $\nu$ is trivial for all
$m_s,m_c,n_s,n_c\in\mathbb Z_2$ precisely when
\begin{align}
h^\vee\gamma_s^2\in2\mathbb Z,
\quad
h^\vee\gamma_c^2\in2\mathbb Z,
\quad
h^\vee(\gamma_s,\gamma_c)\in2\mathbb Z.
\end{align}
For $D_{2r}$,
\begin{align}
h^\vee\gamma_s^2
=
h^\vee\gamma_c^2
=
r(2r-1),
\quad
h^\vee(\gamma_s,\gamma_c)
=
(r-1)(2r-1).
\end{align}
These conditions cannot be satisfied simultaneously. Thus $\nu$ is
nontrivial for $\mathcal C=\mathbb Z_2\times\mathbb Z_2$ for every $r$.

We note that these conditions have an equivalent interpretation in terms of Chern–Simons admissibility. Assuming that $k$ is CS admissible, the level-independent discrete torsion data are unchanged under the fermionic shift $\kappa=k+h^\vee$. Hence $\kappa$ is also CS admissible precisely when
\begin{align}
\frac{h^\vee}{2}\omega^2\in\mathbb Z,
\quad
h^\vee(\mu,\omega)\in\mathbb Z,
\quad
\mu,\omega\in\mathcal C.
\end{align}
These are exactly the conditions for $\nu$ to be trivial. Thus, $\nu$ is trivial precisely when $k$ and $\kappa=k+h^\vee$ are simultaneously CS admissible. In contrast, simultaneous WZW admissibility is automatic, since $h^\vee$ itself is WZW admissible.

The complete classification is summarized in
Table~\ref{table:nu-triviality}.
\begin{table}[h]
\centering
\renewcommand{\arraystretch}{1.4}
\begin{tabular}{c|c|c}
\hline
$\mathfrak g$ & $\mathcal C$ & $\nu\ {\rm trivial?}$\\
\hline
$A_{r-1}$
& $\mathbb Z_N,\ N\mid r$
& $r\ {\rm odd}\ \text{or}\ r/N\ {\rm even}$
\\
\hline
$B_r$
& $\mathbb Z_2$
& ${\rm no}$
\\
\hline
$C_r$
& $\mathbb Z_2$
& $r\equiv0,3\pmod4$
\\
\hline
$D_{2r+1}$
&
$\begin{array}{c}
\mathbb Z_2\\
\mathbb Z_4
\end{array}$
&
$\begin{array}{c}
{\rm yes}\\
r\ {\rm even}
\end{array}$
\\
\hline
$D_{2r}$
&
$\begin{array}{c}
\mathbb Z_2^v\\
\mathbb Z_2^{s,c}\\
\mathbb Z_2\times\mathbb Z_2
\end{array}$
&
$\begin{array}{c}
{\rm yes}\\
r\ {\rm even}\\
{\rm no}
\end{array}$
\\
\hline
$E_6$
& $\mathbb Z_3$
& ${\rm yes}$
\\
\hline
$E_7$
& $\mathbb Z_2$
& ${\rm no}$
\\
\hline
$E_8,F_4,G_2$
& $1$
& ${\rm yes}$
\\
\hline
\end{tabular}
\caption{Triviality of the fermionic phase $\nu$ for the possible
global forms.}
\label{table:nu-triviality}
\end{table}

\section{Quadratic refinements, Arf and relative Rochlin invariants}
\label{apx:arfRochlin}

In this appendix we review the relation between spin structures and
quadratic refinements of the intersection form on the torus, and the
relation between relative Rochlin invariants and fermion Pfaffian
holonomies. The application to the adjoint fermions of the SWZW model
is given in Sec.~\ref{subsec:pfaffian}.

On a torus $T^2$, consider
\begin{align}
H_1(T^2,\mathbb Z_2)
=
\mathbb Z_2 a\oplus\mathbb Z_2 b.
\end{align}
The intersection pairing
$I:H_1(T^2,\mathbb Z_2)\times H_1(T^2,\mathbb Z_2)\to\mathbb Z_2$
satisfies
\begin{align}
I(a,a)=I(b,b)=0,
\qquad
I(a,b)=I(b,a)=1.
\end{align}
A quadratic refinement of $I$ is a function
$q:H_1(T^2,\mathbb Z_2)\to\mathbb Z_2$ satisfying
\begin{align}
q(x+y)
=
q(x)+q(y)+I(x,y).
\end{align}
Spin structures on $T^2$ are in one-to-one correspondence with
quadratic refinements of $I$~\cite{Johnson1980}. A mod-$2$ twist
$s\in H_1(T^2,\mathbb Z_2)$ acts by translating the quadratic
refinement,
\begin{align}
q_s(x)
=
q(x)+I(s,x).
\end{align}
Since $-s=s$ in $H_1(T^2,\mathbb Z_2)$, one has $q_{-s}=q_s$.

The Arf invariant of $q$ is
\begin{align}
{\rm Arf}(q)
=
q(a)q(b)
\in\mathbb Z_2.
\end{align}
For $s=s_a a+s_b b$,
\begin{align}
q_s(a)
=
q(a)+s_b,
\qquad
q_s(b)
=
q(b)+s_a,
\end{align}
and therefore
\begin{align}
{\rm Arf}(q_s)-{\rm Arf}(q)
&=
s_a q(a)+s_b q(b)+s_a s_b
\nonumber\\
&=
q(s)
\pmod 2.
\end{align}
Thus the change of the Arf invariant under a mod-$2$ twist is determined
by the quadratic refinement evaluated on the twisting class.

Through the relative Rochlin invariant, the change of spin structure
determines the corresponding relative Pfaffian holonomy
\cite{LeeMillerWeintraub1988}. We denote the relative Pfaffian holonomy
between the spin structures associated with $q_s$ and $q$ by
\begin{align}
\frac{{\rm Pf}(q_s)}{{\rm Pf}(q)}.
\end{align}
Here the ratio denotes the holonomy of the relative Pfaffian line,
rather than a literal quotient of Pfaffian sections.
Denoting by $D_s$ and $D$ the mapping-torus Dirac operators associated
with the twisted and untwisted spin structures, respectively, the
relative Pfaffian holonomy may equivalently be expressed in terms of
the reduced eta invariant defined in~\eqref{eq:reducedEta}.
In terms of the corresponding
mod-$8$ Rochlin invariant $A(q)$, with
\begin{align}
A(q)=4\,{\rm Arf}(q)\pmod 8,
\end{align}
these equivalent descriptions give
\begin{align}
\frac{{\rm Pf}(q_s)}
     {{\rm Pf}(q)}
&=
\exp\{
- \pi i
\left[
\bar\eta(D_s)-\bar\eta(D)
\right]
\}
\nonumber\\
&=
\exp\{
-\frac{2\pi i}{16}
\left[
A(q_s)-A(q)
\right]
\}
=
(-i)^{q(s)}.
\end{align}
We now specialize to the periodic--periodic (PP) spin structure relevant
for the SWZW partition function. Its quadratic refinement satisfies
\begin{align}
q^{\rm PP}(a)
=
q^{\rm PP}(b)
=
1,
\end{align}
and hence ${\rm Arf}(q^{\rm PP})=1$. Therefore
\begin{align}
{\rm Arf}(q^{\rm PP}_s)
-
{\rm Arf}(q^{\rm PP})
=
s_a+s_b+s_a s_b
=
q^{\rm PP}(s)
\pmod 2.
\end{align}
The corresponding relative Pfaffian holonomy is
\begin{align}
\label{eq:relative-Pfaffian-PP}
\frac{{\rm Pf}(q^{\rm PP}_s)}
     {{\rm Pf}(q^{\rm PP})}
=
(-i)^{q^{\rm PP}(s)}
=
(-i)^{s_a+s_b+s_a s_b}.
\end{align}

This gives the relative Pfaffian phase of the PP spin structure under a
mod-$2$ twist $s\in H_1(T^2,\mathbb Z_2)$. Its application to the root
spaces of the adjoint Majorana--Weyl fermion yields the global phase
appearing in the one-loop determinant, as discussed in
Sec.~\ref{subsec:pfaffian}.

\section{Generalized Siegel--Narain theta functions}
\label{apx:SiegelNarain}

We collect here the conventions and modular transformation properties
of the generalized Siegel--Narain theta functions used in the main
text. They extend the Siegel--Narain theta function introduced in~\cite{Lu_wzwlocalization2026} by including two characteristics,
corresponding respectively to an insertion and a shift of the Narain
lattice.

Let $\Gamma$ be an even lattice of signature $(d_+,d_-)$, with elements
$p=[p_L;p_R]$,
where $p_L$ and $p_R$ lie in Euclidean spaces of dimensions $d_+$ and $d_-$, respectively, equipped with positive-definite inner products $(\, ,\,)$. The Lorentzian pairing decomposes as
\begin{align}
    \langle p,p'\rangle
=
\langle p,p'\rangle_+
+
\langle p,p'\rangle_-
=
(p_L,p'_L)-(p_R,p'_R),
\end{align}
where $\langle p,p'\rangle_+=(p_L,p'_L)$ and $\langle p,p'\rangle_-=-(p_R,p'_R)$.
For $\alpha,\beta\in\Gamma\otimes\mathbb R$, we define
\begin{align}
\mathcal S_\Gamma^{\alpha,\beta}
(\tau,\bar\tau, \theta)
&=
e^{E(\tau, \bar \tau, \theta )}
e^{\pi i\langle\alpha,\beta\rangle}
\sum_{p\in\Gamma+\beta}
e^{-2\pi i\langle p,\alpha\rangle}
 e^{\pi i \tau \langle p, p \rangle_{+}  + \pi i \bar \tau \langle p, p \rangle_{-} +2 \pi i \langle p, \theta \rangle }
\label{eq:generalized-SN}
\end{align}
where $\theta=[\theta_L;\theta_R]\in\Gamma\otimes\mathbb C$, and 
\begin{align}
E(\tau, \bar \tau, \theta )
=
\frac{\pi i}{\tau_2}
\left(
\langle \theta , {\rm Im} \theta \rangle_{+}
-
\langle \theta , {\rm Im} \theta \rangle_{-}
\right).
\label{eq:SN-completion}
\end{align}
For $\alpha=\beta=0$, \eqref{eq:generalized-SN} reduces to the
Siegel--Narain theta function $\mathcal S_\Gamma$ used in~\cite{Lu_wzwlocalization2026}.

The non-holomorphic factor \eqref{eq:SN-completion} compensates the
Gaussian modular anomaly generated by Poisson resummation of the lattice
sum. The completion differs from the one used in~\cite{Gukov:2004id}, which in our conventions takes the form
\begin{align}
F(\tau, \bar \tau, \theta )
=
\frac{\pi}{2\tau_2}
\left(
\langle \theta , \theta \rangle_{+}
-
\langle \theta , \theta \rangle_{-}
\right).
\end{align}
The two are related by
\begin{align}
F(\tau,\bar \tau, \theta)
=
E(\tau, \bar \tau, \theta)
+
H(\tau, \bar \tau,\theta ),
\end{align}
where
\begin{align}
H(\tau,\bar\tau,\theta)
=
\frac{\pi}{2\tau_2}
\left[
\langle\theta,\bar\theta\rangle_+
-
\langle\theta,\bar\theta\rangle_-
\right]= 
\frac{\pi}{2\tau_2}
\left[
(\theta_L,\bar\theta_L)
+
(\theta_R,\bar\theta_R)
\right].
\end{align}
Here $H(\tau,\bar\tau,\theta)$ is modular invariant. Its invariance under $\tau\mapsto\tau+1$, $\theta_L\mapsto\theta_L$, $\theta_R\mapsto\theta_R$ is immediate, while under $S$,
\begin{align}
    \tau\mapsto-\frac1\tau,
\qquad
\theta_L\mapsto\frac{\theta_L}{\tau},
\qquad
\theta_R\mapsto\frac{\theta_R}{\bar\tau},
\end{align}
leads to 
\begin{align}
    \tau_2\mapsto\frac{\tau_2}{|\tau|^2},
\qquad
(\theta_{L,R},\bar\theta_{L,R})
\mapsto
\frac{(\theta_{L,R},\bar\theta_{L,R})}{|\tau|^2},
\end{align}
showing $H$ is also invariant. Hence $E$ and $F$ have identical modular anomalies.

The generalized Siegel--Narain theta function obeys the modular
transformation laws
\begin{align}
\mathcal S^{\alpha,\beta}_\Gamma
(\tau+1,\bar\tau+1,\theta)
&=
\mathcal S^{\alpha-\beta,\beta}_\Gamma
(\tau,\bar\tau,\theta),
\label{eq:SN-T}
\\
\mathcal S^{\alpha,\beta}_\Gamma
\left(
-\frac1\tau,-\frac1{\bar\tau},
\left[\frac{\theta_L}{\tau};\frac{\theta_R}{\bar\tau}\right]
\right)
&=
\frac{
(-i\tau)^{d_+/2}
(i\bar\tau)^{d_-/2}
}{
{\rm Vol}(\Gamma)
}
\mathcal S^{\beta,-\alpha}_{\Gamma^\vee}
(
\tau,\bar\tau,[\theta_L ; \theta_R]
),
\label{eq:SN-S}
\end{align}
where $\Gamma^\vee$ denotes the dual lattice.
For an even unimodular lattice,
$\Gamma^\vee=\Gamma$ and ${\rm Vol}(\Gamma)=1$, so the characteristics
transform simply as
\begin{align}
T:\quad
(\alpha,\beta)
&\longmapsto
(\alpha-\beta,\beta),
\label{eq:SN-characteristics-T}
\\
S:\quad
(\alpha,\beta)
&\longmapsto
(\beta,-\alpha),
\label{eq:SN-characteristics-S}
\end{align}
with modular weight $(d_+/2,d_-/2)$.

\bibliographystyle{myJHEP}
\bibliography{mybibs}

\providecommand{\href}[2]{#2}\begingroup\raggedright\begin{thebibliography}{10}

\bibitem{Witten1983WZW}
E.~Witten, { \it {Nonabelian Bosonization in Two-Dimensions}},   {\rm Commun. Math. Phys.} {\bf 92} (1984) 455--472.

\bibitem{murthyW25}
S.~Murthy and E.~Witten, { \it {Localization of Strings on Group Manifolds}},   {\rm Commun. Math. Phys.} {\bf 407} (2026), no.~7 154, [\href{http://arxiv.org/abs/2506.20028}{{\tt arXiv:2506.20028}}].

\bibitem{Lu_wzwlocalization2026}
Y.~L\"u, { \it Localization and abelianization of strings on group manifolds: The simply connected case},   {\rm To appear} (2026).

\bibitem{DiVecchiaKnizhnikRossi1984SWZW}
P.~Di~Vecchia, V.~G. Knizhnik, J.~L. Petersen, and P.~Rossi, { \it {A Supersymmetric Wess-Zumino Lagrangian in Two-Dimensions}},   {\rm Nucl. Phys. B} {\bf 253} (1985) 701--726.

\bibitem{Wesszumino1971}
J.~Wess and B.~Zumino, { \it {Consequences of anomalous Ward identities}},   {\rm Phys. Lett. B} {\bf 37} (1971) 95--97.

\bibitem{FGK1988}
G.~Felder, K.~Gawedzki, and A.~Kupiainen, { \it Spectra of {Wess}-{Zumino}-{Witten} models with arbitrary simple groups},   {\rm Communications in Mathematical Physics} {\bf 117} (1988), no.~1 127--158.

\bibitem{SchellekensYankielowicz1989Extended}
A.~N. Schellekens and S.~Yankielowicz, { \it {Extended Chiral Algebras and Modular Invariant Partition Functions}},   {\rm Nucl. Phys. B} {\bf 327} (1989) 673--703.

\bibitem{KreuzerSchellekens1993}
M.~Kreuzer and A.~N. Schellekens, { \it {Simple currents versus orbifolds with discrete torsion: A Complete classification}},   {\rm Nucl. Phys. B} {\bf 411} (1994) 97--121, [\href{http://arxiv.org/abs/hep-th/9306145}{{\tt hep-th/9306145}}].

\bibitem{GawedzkiWaldorf2009PWMulGerb}
K.~Gawedzki and K.~Waldorf, { \it {Polyakov-Wiegmann Formula and Multiplicative Gerbes}},   {\rm JHEP} {\bf 09} (2009) 073, [\href{http://arxiv.org/abs/0908.1130}{{\tt arXiv:0908.1130}}].

\bibitem{Witten1985Global}
E.~Witten, { \it {Global gravitational anomalies}},   {\rm Commun. Math. Phys.} {\bf 100} (1985) 197.

\bibitem{LeeMillerWeintraub1988}
R.~Lee, E.~Y. Miller, and S.~H. Weintraub, { \it Rochlin invariants, theta functions and the holonomy of some determinant line bundles},   {\rm J. Reine Angew. Math.} {\bf 392} (1988) 187--218.

\bibitem{Choi:2025MajoWeyl}
C.~Choi, { \it {Global Anomalies in Sigma Models with Majorana--Weyl Fermions}},  \href{http://arxiv.org/abs/2508.14895}{{\tt arXiv:2508.14895}}.

\bibitem{Witten1999Worldsheet}
E.~Witten, { \it {World-Sheet Corrections Via D-Instantons}},   {\rm JHEP} {\bf 02} (2000) 030, [\href{http://arxiv.org/abs/hep-th/9907041}{{\tt hep-th/9907041}}].

\bibitem{SaitoTachikawa:2025GS}
S.~Saito and Y.~Tachikawa, { \it {Cancelling mod-2 anomalies by Green-Schwarz mechanism with $B_{\mu\nu}$}},   {\rm SciPost Phys.} {\bf 19} (2025), no.~1 017, [\href{http://arxiv.org/abs/2411.09223}{{\tt arXiv:2411.09223}}].

\bibitem{BismutFreed1986I}
J.-M. Bismut and D.~S. Freed, { \it The analysis of elliptic families. i. metrics and connections on determinant bundles},   {\rm Comm. Math. Phys.} {\bf 106} (1986), no.~2 159--176.

\bibitem{BismutFreed1986II}
J.~M. Bismut and D.~S. Freed, { \it {The Analysis of Elliptic Families. 2. Dirac Operators, $\eta$ Invariants, and the Holonomy Theorem}},   {\rm Commun. Math. Phys.} {\bf 107} (1986) 103--163.

\bibitem{APS1975Ireducedeta}
M.~F. Atiyah, V.~K. Patodi, and I.~M. Singer, { \it {Spectral asymmetry and Riemannian Geometry 1}},   {\rm Math. Proc. Cambridge Phil. Soc.} {\bf 77} (1975) 43.

\bibitem{DaiFreed1994eta}
X.-z. Dai and D.~S. Freed, { \it {eta invariants and determinant lines}},   {\rm J. Math. Phys.} {\bf 35} (1994) 5155--5194, [\href{http://arxiv.org/abs/hep-th/9405012}{{\tt hep-th/9405012}}]. [Erratum: J.Math.Phys. 42, 2343--2344 (2001)].

\bibitem{Witten2015FermionPI}
E.~Witten, { \it {Fermion Path Integrals And Topological Phases}},   {\rm Rev. Mod. Phys.} {\bf 88} (2016), no.~3 035001, [\href{http://arxiv.org/abs/1508.04715}{{\tt arXiv:1508.04715}}].

\bibitem{Yonekura2016DaiFreed}
K.~Yonekura, { \it {Dai-Freed theorem and topological phases of matter}},   {\rm JHEP} {\bf 09} (2016) 022, [\href{http://arxiv.org/abs/1607.01873}{{\tt arXiv:1607.01873}}].

\bibitem{WittenYonekura2019eta}
E.~Witten and K.~Yonekura, { \it {Anomaly Inflow and the $\eta$-Invariant}},  in  {\rm {The Shoucheng Zhang Memorial Workshop}}, 9, 2019.
\newblock \href{http://arxiv.org/abs/1909.08775}{{\tt arXiv:1909.08775}}.

\bibitem{choiT25}
C.~Choi and L.~A. Takhtajan, { \it {Supersymmetry and trace formulas. Part III. Frenkel trace formula}},   {\rm JHEP} {\bf 06} (2026) 200, [\href{http://arxiv.org/abs/2502.10210}{{\tt arXiv:2502.10210}}].

\bibitem{Koizumi2021DimOne}
S.~Koizumi, { \it {Global anomalies and bordism invariants in one dimension}},   {\rm J. Math. Phys.} {\bf 64} (2023), no.~9 092301, [\href{http://arxiv.org/abs/2111.15254}{{\tt arXiv:2111.15254}}].

\bibitem{FreedVafa87Global}
D.~S. Freed and C.~Vafa, { \it {Global anomalies on orbifolds}},   {\rm Commun. Math. Phys.} {\bf 110} (1987) 349. [Addendum: Commun. Math. Phys. 117, 349 (1988)].

\bibitem{Wendt2001WZW}
R.~Wendt, { \it A symplectic approach to certain functional integrals and partition functions},   {\rm Journal of Geometry and Physics} {\bf 40} (2001), no.~1 65--99.

\bibitem{witten912dYM}
E.~Witten, { \it {On quantum gauge theories in two-dimensions}},   {\rm Commun. Math. Phys.} {\bf 141} (1991) 153--209.

\bibitem{witten922dYMrevisited}
E.~Witten, { \it {Two-dimensional gauge theories revisited}},   {\rm J. Geom. Phys.} {\bf 9} (1992) 303--368, [\href{http://arxiv.org/abs/hep-th/9204083}{{\tt hep-th/9204083}}].

\bibitem{Witten1988CSJones}
E.~Witten, { \it {Quantum Field Theory and the Jones Polynomial}},   {\rm Commun. Math. Phys.} {\bf 121} (1989) 351--399.

\bibitem{elitzurMooreSeibergSchwimmer89}
S.~Elitzur, G.~W. Moore, A.~Schwimmer, and N.~Seiberg, { \it {Remarks on the Canonical Quantization of the Chern-Simons-Witten Theory}},   {\rm Nucl. Phys. B} {\bf 326} (1989) 108--134.

\bibitem{BeasleyWitten2005}
C.~Beasley and E.~Witten, { \it {Non-Abelian localization for Chern-Simons theory}},   {\rm J. Diff. Geom.} {\bf 70} (2005), no.~2 183--323, [\href{http://arxiv.org/abs/hep-th/0503126}{{\tt hep-th/0503126}}].

\bibitem{BlauThompson2006S1bundle}
M.~Blau and G.~Thompson, { \it {Chern-Simons theory on S1-bundles: Abelianisation and q-deformed Yang-Mills theory}},   {\rm JHEP} {\bf 05} (2006) 003, [\href{http://arxiv.org/abs/hep-th/0601068}{{\tt hep-th/0601068}}].

\bibitem{Forste:2003km}
S.~F{\"o}rste and D.~Roggenkamp, { \it {Current-current deformations of conformal field theories, and WZW models}},   {\rm JHEP} {\bf 05} (2003) 071, [\href{http://arxiv.org/abs/hep-th/0304234}{{\tt hep-th/0304234}}].

\bibitem{Kiritsis:1993ju}
E.~Kiritsis, { \it {Exact duality symmetries in CFT and string theory}},   {\rm Nucl. Phys. B} {\bf 405} (1993) 109--142, [\href{http://arxiv.org/abs/hep-th/9302033}{{\tt hep-th/9302033}}].

\bibitem{Gaberdiel:1995mx}
M.~R. Gaberdiel, { \it {Abelian duality in WZW models}},   {\rm Nucl. Phys. B} {\bf 471} (1996) 217--232, [\href{http://arxiv.org/abs/hep-th/9601016}{{\tt hep-th/9601016}}].

\bibitem{Maloney:2020nni}
A.~Maloney and E.~Witten, { \it {Averaging over Narain moduli space}},   {\rm JHEP} {\bf 10} (2020) 187, [\href{http://arxiv.org/abs/2006.04855}{{\tt arXiv:2006.04855}}].

\bibitem{Afkhami-Jeddi:2020ezh}
N.~Afkhami-Jeddi, H.~Cohn, T.~Hartman, and A.~Tajdini, { \it {Free partition functions and an averaged holographic duality}},   {\rm JHEP} {\bf 01} (2021) 130, [\href{http://arxiv.org/abs/2006.04839}{{\tt arXiv:2006.04839}}].

\bibitem{Dong:2021wot}
J.~Dong, T.~Hartman, and Y.~Jiang, { \it {Averaging over moduli in deformed WZW models}},   {\rm JHEP} {\bf 09} (2021) 185, [\href{http://arxiv.org/abs/2105.12594}{{\tt arXiv:2105.12594}}].

\bibitem{Dymarsky:2020qom}
A.~Dymarsky and A.~Shapere, { \it {Quantum stabilizer codes, lattices, and CFTs}},   {\rm JHEP} {\bf 03} (2021) 160, [\href{http://arxiv.org/abs/2009.01244}{{\tt arXiv:2009.01244}}].

\bibitem{GawedzkiReis2002braneGerbe}
K.~Gawedzki and N.~Reis, { \it {WZW branes and gerbes}},   {\rm Rev. Math. Phys.} {\bf 14} (2002) 1281--1334, [\href{http://arxiv.org/abs/hep-th/0205233}{{\tt hep-th/0205233}}].

\bibitem{ChoiHKKusukiO2025BerryBCFT}
Y.~Choi, H.~Ha, D.~Kim, Y.~Kusuki, S.~Ohyama, and S.~Ryu, { \it {Higher structures on boundary conformal manifolds: Higher Berry phase and boundary conformal field theory}},   {\rm Phys. Rev. D} {\bf 113} (2026), no.~10 106005, [\href{http://arxiv.org/abs/2507.12525}{{\tt arXiv:2507.12525}}].

\bibitem{GaberdielGannon2004Dbrane}
M.~R. Gaberdiel and T.~Gannon, { \it {D-brane charges on nonsimply connected groups}},   {\rm JHEP} {\bf 04} (2004) 030, [\href{http://arxiv.org/abs/hep-th/0403011}{{\tt hep-th/0403011}}].

\bibitem{FreedWitten1999AnomalyDbrane}
D.~S. Freed and E.~Witten, { \it {Anomalies in string theory with D-branes}},   {\rm Asian J. Math.} {\bf 3} (1999) 819, [\href{http://arxiv.org/abs/hep-th/9907189}{{\tt hep-th/9907189}}].

\bibitem{Kapustin:1999DbraneBfield}
A.~Kapustin, { \it {D-branes in a topologically nontrivial B field}},   {\rm Adv. Theor. Math. Phys.} {\bf 4} (2000) 127--154, [\href{http://arxiv.org/abs/hep-th/9909089}{{\tt hep-th/9909089}}].

\bibitem{FredenhagenSchomerus2000Ktheory}
S.~Fredenhagen and V.~Schomerus, { \it {Branes on group manifolds, gluon condensates, and twisted K theory}},   {\rm JHEP} {\bf 04} (2001) 007, [\href{http://arxiv.org/abs/hep-th/0012164}{{\tt hep-th/0012164}}].

\bibitem{Gawedzki2004AbelianNonAbelian}
K.~Gawedzki, { \it {Abelian and non-Abelian branes in WZW models and gerbes}},   {\rm Commun. Math. Phys.} {\bf 258} (2005) 23--73, [\href{http://arxiv.org/abs/hep-th/0406072}{{\tt hep-th/0406072}}].

\bibitem{MooreSeiberg1989Taming}
G.~W. Moore and N.~Seiberg, { \it {Taming the Conformal Zoo}},   {\rm Phys. Lett. B} {\bf 220} (1989) 422--430.

\bibitem{Intriligator1989Bonus}
K.~A. Intriligator, { \it {Bonus Symmetry in Conformal Field Theory}},   {\rm Nucl. Phys. B} {\bf 332} (1990) 541--565.

\bibitem{Johnson1980}
D.~Johnson, { \it Spin structures and quadratic forms on surfaces},   {\rm Journal of the London Mathematical Society} {\bf s2-22} (1980), no.~2 365--373.

\bibitem{Gukov:2004id}
S.~Gukov, E.~Martinec, G.~W. Moore, and A.~Strominger, { \it {Chern-Simons gauge theory and the AdS(3) / CFT(2) correspondence}},  in  {\rm {From Fields to Strings: Circumnavigating Theoretical Physics: A Conference in Tribute to Ian Kogan}}, pp.~1606--1647, 3, 2004.
\newblock \href{http://arxiv.org/abs/hep-th/0403225}{{\tt hep-th/0403225}}.

\end{thebibliography}\endgroup

\end{document}